\documentclass[]{aa}  

\usepackage{graphicx}
\usepackage{txfonts}
\usepackage{xcolor}
\usepackage{amsmath}
\usepackage{amsfonts}
\usepackage{natbib}
\usepackage{longtable}
\usepackage{booktabs}
\usepackage{multirow}
\usepackage[amsmath,thmmarks]{ntheorem}
\usepackage[normalem]{ulem}
\usepackage{lscape}             % to rotate a single page table, example in appendix.
\usepackage{placeins}           % useful with \FloatBarrier, to keep 
\theoremseparator{.}
\theorembodyfont{\normalfont} 
{\bfseries}{\itshape}

\defcitealias{expansion_method}{the method paper}

\usepackage[normalem]{ulem}
\usepackage{hyperref}
\hypersetup{
    colorlinks=true,
    linkcolor=blue,
    filecolor=magenta,      
    urlcolor=cyan,
    citecolor=teal,
    pdftitle={BALYSA: I},
    }
\begin{document}

   \title{Bayesian ages of local young stellar associations}
   \subtitle{I. Through the expansion rate method}

   \author{J. Olivares\inst{1} \and N. Miret-Roig\inst{2} \and P.A.B. Galli\inst{3} \and H. Bouy\inst{4}}
   
   \institute{
   		Departamento de Inteligencia Artificial, Universidad Nacional de Educación a Distancia (UNED), c/Juan del Rosal 16, E-28040, Madrid, Spain. 
   		\email{jolivares@dia.uned.es}
		\and
		University of Vienna, Department of Astrophysics, Türkenschanzstraße 17, 1180 Wien, Austria
		\and
		Instituto de Astronomia, Geofísica e Ciências Atmosféricas, Universidade de São Paulo, Rua do Matão, 1226, Cidade Universitária, 05508-090 São Paulo-SP, Brazil
		\and
		Laboratoire d'astrophysique de Bordeaux, Univ. Bordeaux, 
   		CNRS, B18N, allée Geoffroy Saint-Hilaire, 33615 Pessac, France.
        }

   \date{}

% \abstract{}{}{}{}{}
% 5 {} token are mandatory
 
  \abstract
   {Local young stellar associations (LYSAs <50 Myr and <150 pc) are important laboratories to test predictions from star-formation theories. Estimating their ages through various dating techniques with minimal biases is thus of paramount importance.}
   {We aim at determining the ages of LYSAs with the expansion rate dating technique.} 
   {We estimate the ages of the LYSAs using literature membership lists, publicly available data (astrometry and radial velocities), and a recent open-source Bayesian code that implements the expansion rate method. This code in combination with simple statistical assumptions allow us to decontaminate, identify possible substructures or populations, and estimate expansion ages.}
   {We derive the largest and most methodological homogeneous set of ages of LYSAs. We rediscover three and discover four associations hidden within the literature membership lists of the classical ones.}
   {The expansion ages we report here are compatible with literature age estimates. Moreover, our analysis shows that previous age tensions can be explained, in most cases, by the presence of unidentified populations or substructures.}


   \maketitle
%
%________________________________________________________________
\section{Introduction}
\label{introduction}

Accurate and precise age determinations of stellar associations are fundamental stepping stones for the validation of the current theories of the formation of stars and planets and their dynamical evolution. Yet, inferring these ages is a difficult task due to the inherent complexities associated with the data availability and quality, and the often inconsistent results obtained through different dating techniques \citep[see, for example][]{2022A&A...664A..70G,1999ApJ...522L..53B,1995AJ....109..298S}.

Kinematic ages and nuclear ages are the two most common dating techniques applied to stellar associations. On one hand, the dynamical or kinematic dating techniques \citep[e.g., expansion rate and traceback, see for example,][and references therein]{2024A&A...689A..11G,2023MNRAS.520.6245G,2024MNRAS.533..705W,2022A&A...667A.163M,1997MNRAS.285..479B} take advantage of the proximity and youth of stellar associations to determine their age based on its kinematic properties. The underlying assumption of these techniques is that stellar associations were more spatially concentrated at their birth than what they are today, and that the elapsed time can be inferred based on the current positions and velocities of its members (while the traceback technique integrates these back in time under a Galactic potential, the expansion rate assumes that they resulted from ballistic trajectories). On the other hand, the nuclear dating techniques \citep[e.g., isochrone-fitting and Lithium depletion boundary, hereafter LDB, see for example,][and references therein]{2023A&A...678A..71R,2023MNRAS.523..802J,2022A&A...664A..70G,2006ApJ...645.1436V} derive ages by fitting the predictions of stellar interior and stellar atmospheric models, which are assumed to be correct and unbiased, to the ensemble of observed photometric or spectroscopic properties of the system's members.

In this series of articles, we will infer the ages of the local young stellar associations (LYSA) with dynamical and nuclear dating techniques under a Bayesian paradigm. In this first article, we aim at inferring the ages of LYSA with the expansion rate dating technique \citep[for an application of this method see, for example,][]{2014MNRAS.445.2169M}, particularly with the Bayesian age estimator recently developed by \citet[][hereafter \citetalias{expansion_method}]{expansion_method}. The expansion rate of a stellar system measures the average rate of change in the system's velocity with respect to the distance to its centre \citep[see, for example, the section "The concept of linear expansion" in][]{1964ARA&A...2..213B}. Thus, in the expansion rate dating method, this spatial rate of expansion is inverted to obtain an estimate of the stellar system's age \citepalias[see Eq. 1 of][]{expansion_method}. We notice that this method makes the underlying assumption that the stellar system is expanding, that this expansion originated at the system's birth, and that it has remained constant since then (more below).

The expansion of young stellar systems was hypothesised long ago by \citet{1955Obs....75...72A,1954LIACo...5..293A} but remained elusive due to examples of detections \citep{2015ApJ...812..131K}, non-detections \citep{2017ApJ...834..139D,2016MNRAS.460.2593W}, and partial detections through runaway members \citep{2013AJ....146..106O,2005ApJ...627L..61P}. It was not until recently and thanks to exquisite astrometry of the \textit{Gaia} data, that the expansion of young stellar systems has been statistically confirmed in relatively large samples \citep[e.g.,][]{2024MNRAS.533..705W,2019ApJ...870...32K}. The expansion of young stellar associations may result from a variety of phenomena, such as changes in their gravitational potential due to gas expulsion \citep[e.g.,][]{1980ApJ...235..986H,1978A&A....70...57T} or tidal shocks by surrounding stellar systems or gas clouds \citep{2012MNRAS.419..841K}. Despite its origin, once detected, the expansion rate can be used to compute the time elapsed between the event that originated the expansion and the present day. Moreover, if the expansion started at birth time and remained constant since, then the age of the stellar system can be estimated by inverting its expansion rate. Throughout this work, we will assume that, whenever expansion is detected, it was imprinted at birth, and has remained constant since then  \citepalias[see Assumptions 1, 2, 3 and Sect. 6.3 of][]{expansion_method}.

The rest of this work is organised as follows. In Sect. \ref{data}, we present the data of selected LYSAs that fall within the applicability domain of the Bayesian expansion rate method. In Sect. \ref{methods}, we briefly describe the methods we will use. Then, in Sect. \ref{results}, we show the results obtained by the Bayesian age estimator, and in Sect. \ref{discussion}, we discuss them. Finally, in Sect. \ref{conclusions}, we present our conclusions.

\section{Data}
\label{data}

The applicability domain of the Bayesian expansion rate age estimator (see Sect. \ref{methods}) when applied to stellar associations covers from $\sim$10 Myr to $\sim$40 Myr in age and up to $\sim$150 pc in distance \citepalias[see Sect. 5.1 of][]{expansion_method}. Thus, in this section, we first select the stellar association falling within this domain. Then, we compile their lists of members from the literature together with their astrometry and radial velocity (RV) data.

\subsection{Stellar associations within the applicability domain}
\label{data:stellar_associations}

Our objective is the determination of ages for the local young stellar associations up to 150 pc away and up to 40 Myr old. Thus, we exclude open clusters and star-forming regions within these intervals. On one hand, open clusters are expected to be gravitationally bound to some extent \citep[see Sect. 1.2 of][]{2019ARA&A..57..227K}, and thus, we do not expect them to show expansion rates. To the best of our knowledge, there is no evidence in the literature for the expansion of open clusters' central regions. On the other hand, star-forming regions are expected to be expanding, however their elevated complexity resulting from the entanglement of their subgroups and partial embedding \citep[e.g.,][]{2023A&A...677A..59R,2023A&A...671A...1O,2022A&A...667A.163M,2019A&A...630A.137G} precludes the direct application of our method. In the future, we will estimate the expansion ages of star-forming regions on a region-by-region basis.

We compiled a list of nearby young stellar associations from the works of \cite{2017AJ....153...95R,2018ApJ...856...23G,2018ApJ...860...43G, 2018ApJ...862..138G,2019MNRAS.486.3434L}, and \cite{2022ApJ...939...94M}. We gather ages and distances from Table 1 of \citet{2018ApJ...860...43G} and Table 2 of \citet{2019MNRAS.486.3434L}. Then, we filter literature lists of stellar associations and end up with the following ones (see Table \ref{table:summary} for details): \object{$\epsilon$-Chamaeleontis} (EPCHA), \object{$\eta$-Chamaeleontis} (ETCHA), \object{TW Hydrae} (TWA), \object{118 Tau} (118TAU),  \object{32 Orionis} (THOR), \object{$\beta$-Pictoris} (BPIC), \object{Octans} (OCT), \object{Columba} (COL), \object{Carina} (CAR), and \object{Tucana-Horologium} (THA).

The previous associations are briefly summarised as follows. EPCHA and ETCHA are two young stellar associations located on the outskirts of the Scorpius-Centaurus region, with similar ages and distances to the Sun: $\sim$9 Myr and $\sim$100 pc, respectively \citep[see for example,][]{2023A&A...678A..71R}. Due to their similar space velocities, they have been recognised as related between themselves and to the background star-forming regions of Chamaeleon I and II \citep{2024RAA....24e5004B,2021A&A...646A..46G}. Although their age is at the lower limit of the applicability domain of our method, we nonetheless include them in our analysis. TWA is one of the youngest and closest to the Sun associations \citep[40-100 pc, $\sim$10 Myr,][]{2023AJ....165..269L}, which makes it a fundamental system for the study of the solar neighbourhood's star-formation history. THOR and 118TAU are two nearby groups located at approximately 100 pc and 110 pc, respectively, and with ages between 17 and 25 Myr \citep{2022AJ....164..151L,2022A&A...664A..70G,2021ApJ...917...23K,2017MNRAS.468.1198B,2016ApJ...820...32B,2007IAUS..237..442M}. BPIC, due to its proximity and young age \citep[40-50 pc and 18-24 Myr, e.g.,][]{2024MNRAS.tmp...30L,2022A&A...664A..70G,2020A&A...642A.179M} is a benchmark for the analysis of the formation and evolution of stellar systems and the search of exoplanets. OCT is a relatively young, $\sim$34 Myr, sparse stellar association located at $\sim$150 pc that extends over 40 pc \citep[e.g][]{2024A&A...689A..11G}. COL is a young association \citep[][34-45 Myr]{2024AJ....168..159L,2015MNRAS.454..593B} located $\sim$60 pc away from the Sun. CAR is a nearby stellar association located $\sim$90 pc away and with age estimates of 30-45 Myr \cite[see, Table 2 of][]{2019MNRAS.486.3434L}. Finally, THA is a nearby association located at $\sim$46 pc \citep{2018ApJ...856...23G} with age estimates varying from 27 to 53 Myr \citep[see Sect. 1 of ][]{2024A&A...689A..11G}.

\subsection{Membership lists}
\label{data:membership}

For each stellar association in our list, we gathered its literature members from the following works on stellar association surveys: LACEwING \citep{2017AJ....153...95R}, BANYAN \citep{2018ApJ...856...23G,2018ApJ...860...43G,2018ApJ...862..138G}, \citet{2019MNRAS.486.3434L}, SPYGLASS \citep{2021ApJ...917...23K}, and \cite{2022ApJ...939...94M}. The LACEwING list corresponds to Table 3 from  \citet{2017AJ....153...95R}. Although \citet{2020AJ....159..166U} reanalysed the LACEwING catalogue of stellar association members obtained with \textit{Gaia} DR2 data, here we work with the \textit{Gaia} DR3 data of the original lists by \citet{2017AJ....153...95R}. The BANYAN list consists of the union of Tables 5 and 7 of \citealt{2018ApJ...856...23G}, Table 4 of \citealt{2018ApJ...860...43G}, and Table 2 of \citealt{2018ApJ...862..138G}. The membership lists of \citet{2019MNRAS.486.3434L} are those presented in their Appendix B. However, we do not use the membership lists of \citet{2019MNRAS.489.2189L} because the machine-learning algorithms they use result in the mixing of several stellar associations. Finally, the lists of SPYGLASS and \citet{2022ApJ...939...94M} correspond to their Table 1.

We complement the previous lists of members with the following works targeting specific associations. In EPCHA, with those of SigMA \citep{2023A&A...677A..59R} and \cite{2023ApJS..265...12Q}. In ETCHA, with those of \cite{2024A&A...686A..42H} and SigMA \citep{2023A&A...677A..59R}. In TWA, those by \citet{2023AJ....165..269L} and \cite{2025A&A...694A..60M}, which share the same members but differ in the RV data. In THOR with those by \citet[][their final list]{2023A&A...678A..75Z} and \citet{2024A&A...686A..42H}. In BPIC, the works by  \citet{2024AJ....168..159L}, \citet{2023ApJ...946....6C}, \citet{2020A&A...642A.179M} and \citet{2019MNRAS.489.3625C}. In OCT, with that by \citet{2024A&A...689A..11G}. In COL with that by \citet{2024AJ....168..159L}. In CAR with those by  \citet{2021MNRAS.500.5552B} and \citet{2024AJ....168..159L}. In THA with those by \cite{2023ApJ...945..114P}, \cite{2023MNRAS.520.6245G}, and \citet{2024AJ....168..159L}. 

Finally, we also analyse each association with a list of members resulting from the union  of all its literature lists, excluding duplicated sources, that we call the UNION list. However, we notice that the membership lists by \citet{2023AJ....165..269L} and \citet{2024AJ....168..159L} for 32 Orionis and Car-Ext, respectively, are extremely numerous compared to the rest of the literature lists. For this reason, we exclude these lists from the UNION ones and instead, we analyse them independently.

\subsection{Astrometric data}
\label{data:astrometry}

For each stellar association in our list, we cross-match (using a radius of 10\arcsec) its literature lists of members (above) with the \textit{Gaia} DR3 data. From this latter, we recover sky positions, proper motions, parallax, radial velocity, uncertainties, and correlations. In addition, we also collected the source's photometry and renormalised unit weight error (RUWE).

\subsection{Radial velocities}
\label{data:radial_velocities}

We use the RV data from several sources. In those specific works from the literature where the authors provide RV measurements, we use them to perform independent analyses on those data. We call these RVs catalogues the "original" ones.

In addition, we compile a catalogue with publicly available RV measurements, which we call the GLASSS one. It includes RV data from \textit{Gaia} DR3, the Large Sky Area Multi-Object Fiber Spectroscopic Telescope DR10 \citep[LAMOST,][]{2012RAA....12.1197C,2012RAA....12..723Z}, the Apache Point Observatory Galaxy Evolution Experiment DR17 \citep[APOGEE,][]{2022ApJS..259...35A}, the Survey of Surveys DR1 \citep[SoS,][]{2022A&A...659A..95T}, and SIMBAD \citep{2000A&AS..143....9W}. The radial velocities from these surveys were not mixed, but simply  selected according to the following priority: APOGEE, LAMOST, \textit{Gaia}, SoS, and SIMBAD. If neither of these surveys has an RV measurement for the source, we set its value as missing.

\section{Methods}
\label{methods}

In this section, we present the methods used to infer the expansion ages of the LYSA selected in Sect. \ref{data}. Given that the core of our Bayesian methodology is the \textit{Kalkayotl} code \citep{2025A&A...693A..12O,2020A&A...644A...7O}, we start by briefly describing it. Then, we dedicate two subsections to explain the procedures we follow to clean the input list of members from possible contaminants and outliers. First, we apply general filters to remove the most clear outliers in the observed space. Then, we apply a dedicated method to further remove contaminants in the phase space. Afterwards, we describe the method that we use to disentangle phase-space substructures with stellar associations, without this disentanglement, the age estimates could be biased. Finally, we end this section with a brief description of the Bayesian expansion rate age estimator developed in \citetalias{expansion_method}. It is important to notice that throughout this work we will undertake the assumptions listed in Appendix A of  \citetalias{expansion_method}.

\subsection{Review of \textit{Kalkayotl}}
\label{methods:kalkayotl}

\textit{Kalkayotl} is a flexible Bayesian hierarchical framework that allows its users to create models of stellar systems in which the source-level parameters could represent individual stellar distances (1D version), positions (3D version), or positions plus velocities (6D version), and global-level parameters could represent the system's distance, location, velocity and its corresponding dispersions. Furthermore, other global parameters can be inferred, such as contraction, expansion, or rotation. 

The code was specifically designed to take as input the \textit{Gaia} astrometry and possibly missing radial velocities of the stellar system's members. As output, it delivers the joint posterior distribution of the source-level and population-level parameters. It offers a flexible environment where different dimensionalities, statistical distributions, reference systems, and velocity models can be used to construct the desired stellar system's model. In addition, it delivers several diagnostics and summaries of the inferred posterior distribution.

In this work, we only use \textit{Kalkayotl}'s 6D version with parameters inferred in the Galactic reference system with joint and linear velocity fields. As statistical distributions, we only work with the Gaussian one or with Gaussian mixture models (GMM) with two components. 

\subsection{Pre-processing of outliers and unresolved binaries}
\label{methods:outliers}

The presence of outliers, contaminants, and binary stars in the LYSA list of members could bias their parameter estimates, particularly their age. Thus, to minimise these possible biases, we remove as many outliers in the observed space as possible with the following filtering criteria. 

Concerning binary stars, we assume that the most important source of biases corresponds to the RV of unresolved binaries. Therefore, we mask as missing the RV of possibly unresolved binaries, that is, of those sources with RUWE > 1.4. Although the astrometry of unresolved binaries could also be biased, we assume that the observed astrometry corresponds to that of the centre of mass of the binary system \citepalias[see Assumption  5 of ][]{expansion_method}.

Concerning outliers, we consider as such, and thus drop from the input list, those sources with parallax $\geq\!3\sigma_{\varpi}$ and RV $\geq\!2\sigma_{rv}$, with  $\sigma_{\varpi}$ and $\sigma_{rv}$ the standard deviation of the LYSA's observed parallax and RV, respectively. We use a more stringent criterion in the RV than in parallax because we expect stellar associations to be more tightly concentrated in velocity space than in distance. 

Finally, we set a minimum value for the RV uncertainty at $0.01\rm{km\,s^{-1}}$ (i.e. the uncertainty is never smaller than this value). This minimum ensures a fast convergence of \textit{Kalkayotl}'s posterior sampling algorithm without impacting the inferred source-level parameters. 

\subsection{Decontamination algorithm}
\label{methods:decontamination}

Decontaminating the input list of members from a LYSA demands a compromise between completeness and purity. On one hand, contaminants can bias the results towards more random motions, less significant ordered kinematic patterns, and thus, towards older ages. On the other hand, an excessive trimming of the input list of members reduces its statistical significance and thus, diminishes the detectability of the kinematics patterns. Therefore, we present an iterative decontamination algorithm aiming at removing the less sources as possible while rejecting the majority of contaminants. The steps of this decontamination algorithm are the following.

We start by applying \textit{Kalkayotl}'s field-Gaussian mixture decontamination model (FGMM) \citep[see Sect. 2.4.5 of][]{2025A&A...693A..12O}, which is a GMM with two components: one for the field and another for the stellar system. This field-Gaussian distribution is concentric with the Gaussian one of the stellar system and has a diagonal covariance matrix with entries fixed to user-defined values. Here, we set the field standard deviations in positions and velocities to 20 pc and 5 $\rm{km\,s^{-1}}$, which correspond to the upper end values of LYSAs \citep[see Table 1 of][]{2018ApJ...856...23G}. The FGMM returns as output the posterior distribution of the stellar system parameters, the source parameters, and the classification of each source in the input into either the field population or the stellar system. We discard from the list of members those classified by the FGMM as the field population. We iteratively apply this FGMM until the inferred weight of the field population goes less than 10\%. This iterative approach ensures that possible contaminants that escaped from the first classification are removed in subsequent classifications. The 10\% field-weight threshold was heuristically set for these sparse systems. 

Then, we remove possibly unidentified outliers as follows. We apply \textit{Kalkayotl}'s Gaussian model with joint velocity to the sample of members resulting from the previous step. We use the Gaussian with joint velocity model due to its good convergence properties and simplicity in inferring the members' positions and velocities \citep[see Sect. 3.3 of][]{2025A&A...693A..12O}. Following \citet{2023MNRAS.520.6245G} and \citet{2020A&A...642A.179M}, we compute, for each member, its Mahalanobis distance to the stellar association velocity centre with the minimum covariance determinant method \citep[MCD,][]{Rousseeuw1999}, and sort members according to this distance. This filtering method has been successfully applied by our group to remove kinematic outliers and interlopers \citep{2023MNRAS.520.6245G,2020A&A...642A.179M}. Here, we keep all members with Mahalanobis distances of less than 10 $\sigma$, which is a highly conservative criterion. The resulting clean list of members is then used to search for substructures, identify the association's kinematic patterns, and infer its expansion age.

\subsection{Identification of phase-space substructures and kinematic patterns}
\label{methods:substructures}

We search for phase-space substructures within each LYSA using GMMs with two components (see Sect. \ref{methods:kalkayotl}). We start our exploration assuming, as the null hypothesis, that the UNION list of members of each of LYSA (see Sect. \ref{data}) contains only one population and that it can be described with only one Gaussian distribution. As the alternative hypothesis, we assume that the LYSA's UNION list contains only two phase-space substructures, each of them Gaussian distributed, and both can be described by a two-component GMM. Unless specified otherwise, we work with the UNION list of members because, by definition, it has the largest number of sources.  

For the inference of the GMM's parameters, we use the following prior distributions. In the weights parameters, we use a Dirichlet distribution with $a=[1,1]$ (equivalent to a uniform distribution) when there is no prior identification of substructures, or an informed value when substructures were previously identified. The rest of prior distributions and hyper-parameters were set to their default values \citepalias[see Sect. 3.2.2 of][]{expansion_method}.

We adopt the following criteria to reject the null hypothesis. We assume that the association is composed of two populations whenever the Gaussian distributions inferred by the GMM have non-negligible weights (i.e. both weights are larger than 5\%) and are mutually exclusive at the 2$\sigma$-level. This last criterion is equivalent to Mahalanobis distances $\mathcal{M}$ between the two components, say A and B, larger than 2, this is $\mathcal{M}_{AB}>2$ and $\mathcal{M}_{BA}>2$. If the inferred Gaussian distributions have non-negligible weights but are not mutually exclusive at the 2$\sigma$-level, then we say that the LYSA is composed of a core (i.e. the most compact) and halo (i.e. the most extended) populations.

Whenever the alternative two-populations hypothesis proves to be true, we apply the same previous procedure to each of the identified substructures. This exploration allows us to identify more complex substructures with the same underlying assumptions. This hierarchical tree has already been proved in star-forming regions \citep[see][]{2023A&A...671A...1O} and open clusters \citep[see][]{2023A&A...675A..28O}.

The kinematic patterns of expansion, contraction, and rotation are inferred using \textit{Kalkayotl}'s Gaussian family with linear velocity field model, which is described in Sect. 2.3.2 of \citet{2025A&A...693A..12O}. We apply this model to each LYSA and its identified substructures. Finally, we report the detection of such patterns whenever their statistical significance exceeds the $2\sigma$ level.

\subsection{Bayesian expansion rate dating method}
\label{methods:bayesian_ages}

As a methodology to estimate the ages of LYSAs, we use the  Bayesian expansion rate dating method recently introduced in \citetalias{expansion_method}. This method assumes that the age of a stellar system can be estimated by inverting its present-day expansion rate, that this later started at the system's birth time, and that it has remained constant since then (see the aforementioned work's Assumptions). Briefly, the method is built within \textit{Kalkayotl} and upon an improved version of \cite{2000A&A...356.1119L} linear velocity field model in which the diagonal entries of the linear velocity tensor correspond to the expansion rate components. These components are drawn, in a Bayesian hierarchical fashion, from a unique expansion rate that in turn is sampled and inverted from the age distribution. For more details, see Sect. 3 of  \citetalias{expansion_method}.

The extensive validation of hte Bayesian method on realistic synthetic datasets of stellar associations and star-forming regions shows that it offers several advantages over the traditional frequentist approaches of the expansion rate dating techniques at the cost of being more computationally expensive \citepalias[see Sect. 6.1 of][]{expansion_method}. Among these advantages, we consider its ability to work with missing RV data, its high credibility (i.e. the true age value is always contained within the 95\% high density interval of the posterior distribution), low error (<10\%) and low uncertainty (<20\%) as the most important ones. 

We notice that, contrary to other frequentist age estimators, the Bayesian one benefits from larger samples of stellar system members despite their fraction of missing radial velocities. For this reason, in this work we focus on compiling the most complete lists of LYSA's member in spite of their possibly large fractions of missing radial velocities. 

The Bayesian expansion rate method requires the specification of prior distributions for all its population-level parameters (the source-level parameters are sampled from the population-level distribution in a hierarchical fashion). We use the default and non-isotropic configurations of prior distributions and hyper parameters, which are described in Sect. 3.2.2 of  \citetalias{expansion_method}. As recommended in the aforementioned work, we inform the prior distribution of the velocity dispersion using, as a location hyper-parameter, the velocity dispersion value inferred with the joint velocity model on the same dataset. 

Finally, we establish age prior distribution using the available information in the literature and following the recommendations given in Sect. 6.2 of  \citetalias{expansion_method}. In particular, we establish age prior distributions with a mean and a standard deviation under the weakly informative paradigm and repeat the analysis with updated prior distributions whenever we detect hints of posterior biases due to the prior. Appendix \ref{appendix:prior} contains the details of our chosen age prior distribution for each LYSA, as well as the literature works that provided the a priori information. 

\section{Results}
\label{results}

In this Section, we present the results obtained after applying our  methods (see Sect. \ref{methods}) to the selected LYSAs (see Sect. \ref{data}). First, we start with an overview of our results in all associations. Then, for each LYSA, we detail its identified substructures, kinematic patterns, and expansion ages as inferred from the literature membership list.

In Appendix \ref{appendix:tables}, we provide tables with the complete properties of the input lists of members and summary statistics of the model parameters posterior distributions for each LYSA, identified substructure, membership list, and RV origin. As a summary of our results, Table \ref{table:summary} provides the ages inferred for each of LYSA, as well as those  others recovered or discovered within the literature membership lists. 

\begin{table*}[ht!]
\caption{Bayesian expansion ages of local young stellar associations.}
\label{table:summary}
\centering
\resizebox{0.8\textwidth}{!}
{
\begin{tabular}{lllcccr}
\toprule
 &  & \multicolumn{2}{c}{RV} & Members & Distance & Age \\
 &  & origin & coverage & final &  & $\mu\pm\sigma$ \\
 &  &  & [\%] &  & [pc] & [Myr] \\
Association & Membership &  &  &  &  &  \\
\midrule
EPCHA & SigMA & GLASSS & 37.8 & 37 & 102 & $6.4\pm2.1$ \\
MuscaFG & UNION-A & GLASSS & 53.8 & 39 & 101 & $10.4\pm1.4$ \\
ETCHA & UNION & GLASSS & 60.0 & 20 & 98 & $7.6\pm3.4$ \\
\multirow[t]{2}{*}{TWA} & L23-A & GLASSS & 58.5 & 41 & 63 & $10.2\pm1.0$ \\
 & L23-B & GLASSS & 53.3 & 15 & 65 & $8.5\pm1.3$ \\
118TAU & UNIONS+L22-A & GLASSS & 60.0 & 35 & 107 & $24.0\pm4.5$ \\
THOR & UNIONS+L22-B & GLASSS & 66.7 & 45 & 101 & $29.8\pm4.9$ \\
Háap & UNIONS+L22-C & GLASSS & 55.6 & 63 & 90 & $25.9\pm4.6$ \\
BPIC & Miret-Roig+2020 & original & 73.1 & 26 & 42 & $23.4\pm4.8$ \\
Balaam & Luhman2024-B & original & 47.6 & 63 & 63 & $19.3\pm4.7$ \\
OCT & UNION-A & GLASSS & 81.2 & 69 & 150 & $30.4\pm4.6$ \\
HSC2597 & UNION-C & GLASSS & 41.2 & 17 & 138 & $39.6\pm6.9$ \\
COL & UNION-A & GLASSS & 59.2 & 125 & 71 & $33.0\pm9.2$ \\
OMAU & UNION-C & GLASSS & 46.7 & 15 & 36 & $40.1\pm5.9$ \\
CAR & Luhman2024-AB & GLASSS & 44.8 & 259 & 112 & $45.4\pm9.2$ \\
Platais8 & Luhman2024-AA & GLASSS & 42.0 & 362 & 133 & $28.3\pm10.9$ \\
Nal & Luhman2024-BA & GLASSS & 11.4 & 35 & 183 & $16.1\pm5.5$ \\
Chem & Luhman2024-BB & GLASSS & 30.3 & 132 & 168 & $22.2\pm2.1$ \\
\multirow[t]{2}{*}{THA} & Lee+2019-A & GLASSS & 58.4 & 161 & 42 & $32.9\pm6.3$ \\
 & Lee+2019-B & GLASSS & 54.1 & 37 & 43 & $38.6\pm5.4$ \\
\bottomrule
\end{tabular}
}
\tablefoot{
The columns in this table correspond, from left to right, to the association's abbreviated name, the selected membership's list (the reasons for this selection will be described at each LYSA's section), the RV origin and percentage of coverage (out of the final members), the final number of members (after the decontamination algorithm), the association's distance, and the inferred Bayesian expansion age. For simplicity, the age is reported as the mean and standard deviation of the inferred age's posterior distribution. Its 95\% HDI (high density interval) is reported in Table \ref{table:ages}.
}
\end{table*}

\subsection{Overview}
\label{results:overview}

The application of our methods in the literature membership lists showed the following interesting patterns. First, as expected, the pre-\textit{Gaia} membership lists from the literature have larger fractions of contaminants than those of the \textit{Gaia} era. For example, in OCT, the LACEwING and BANYAN lists reach contamination levels of up to 43\% and 45\%, respectively. Second, hidden within the literature membership lists, there are several substructures and populations (see Sect. \ref{methods:substructures} for a disambiguation between these terms). Although the majority of the recovered populations and substructures were already discovered in recent literature works, we discovered four new nearby associations. Finally, as a by-product of our analysis, our methodology delivers kinematic parameters of expansion, contraction and rotation. This is the first time that such a large compilation of these detections is reported.

\subsection{$\epsilon-$ and $\eta-$ Chamaeleontis}
\label{results:EPCHA_and_ETCHA} 

We explore the possibility that the UNION lists of EPCHA and ETCHA hosted phase-space structures. To test this hypothesis, we fitted GMM with two components to both membership lists. In the case of ETCHA, the second component was inferred with a negligible weight, thus indicating that the alternative hypothesis was rejected. On the contrary, in the case of EPCHA, the two fitted Gaussian components, that we call UNION-A and UNION-B, have non-negligible weights of 0.63 and 0.37, and Mahalanobis distances $\mathcal{M}_{AB}=3.9$ and $\mathcal{M}_{BA}=3.3$. Therefore, we reject the null hypothesis with a confidence level >99.8\%. Furthermore, we tested the hypothesis that EPCHA may host three populations by fitting a GMM with three components. However, this model resulted in a negligible weight for the third component, which allowed us to conclude that EPCHA hosts two populations. After cross-matching the members of these populations with recent literature censuses, we found that UNION-A corresponds to the Musca-Foreground group identified and described in  \citet{2023A&A...677A..59R} and \citet{2023A&A...678A..71R}. We identified the remaining group with the classical EPCHA association. 

Concerning lists of members, in EPCHA the contamination and RV coverage reach values of up to 36\% in LACEwING and 70\% in \cite{2023ApJS..265...12Q}. In ETCHA, the contamination reaches 41\% in the UNION list and a 61\% RV coverage in the list by \cite{2024A&A...686A..42H}.

Concerning the kinematic patterns, in ETCHA we were unable to detect any kinematic pattern at a $2\sigma$ level. In EPCHA, we  detect expansion along the X and Z directions, but mostly in Z in almost all membership lists. 

Concerning the expansion ages, we observe the following. In EPCHA, our estimates are all compatible (within the 95\% HDI) and consistent, within the uncertainties, except for that obtained from the list by SigMA, which is significantly younger but still compatible with the ages inferred from the UNION and \cite{2023ApJS..265...12Q} lists. In ETCHA, all our age estimates are compatible and consistent.

Finally, from all the age estimators and membership lists, we choose to report the following ones. In the case of EPCHA, we report the age estimate resulting from the SigMA list because it is the most populous one and the one providing the most significant detections of expansion both in X and Z directions.  In the case of ETCHA, we report the age estimate resulting from the UNION list. Moreover, we also report the age of the MuscaFG group from the UNION-A list of EPCHA.

\subsection{TW Hydrae}
\label{results:TWA}

In TWA, we test the alternative hypothesis of two populations in the UNION, \citet{2023AJ....165..269L}, and \citet{2025A&A...694A..60M} membership lists. We expand our search to the latter two lists because, recently, \citet{2025A&A...694A..60M} find two substructures within their list of members, which is the same as that of \citet{2023AJ....165..269L} except for the RV measurements. 

On one hand, the results on the membership lists by \citet{2023AJ....165..269L} and \citet{2025A&A...694A..60M} with their original RV measurements produce GMM with non-negligible weights in the second components although with Mahalanobis distances that are not mutually exclusive at the 2$\sigma$ level criterium. In the membership list by \citet{2023AJ....165..269L}, the Mahalanobis distances are $\mathcal{M}_{AB}=0.9$ and $\mathcal{M}_{BA}=1.8$, while in that by \citet{2025A&A...694A..60M}, the distances were $\mathcal{M}_{AB}=1.5$ and $\mathcal{M}_{BA}=2.35$. Thus, the data from these lists do not allow us to reject the null hypothesis of a single population.

On the other hand, in the UNION and \citet{2023AJ....165..269L} membership lists with GLASSS radial velocities, our substructures identification method found that the null hypothesis of a single population was rejected with high confidence levels (>99\%). In the UNION list, the GMM with two components, which we call UNION-A and UNION-B, have non-negligible weights of 0.74 and 0.26, respectively, with Mahalanobis distances of $\rm{\mathcal{M}_{AB}=3.25}$ and $\rm{\mathcal{M}_{BA}=4.65}$. In the list by \citet{2023AJ....165..269L}, the GMM with two components, which we call L23-A and L23-B, was inferred with non-negligible weights of 0.64 and 0.36, respectively, and their Mahalanobis distances are $\rm{\mathcal{M}_{AB}=3.40}$ and $\mathcal{M}_{BA}=3.45$. Moreover, the groups A and B identified from these two membership lists share most of their members. The 42 members of L23-A are shared with UNION-A and none with UNION-B, while out of the 15 members of L23-B, 6 are shared with UNION-B and only one with UNION-A. Thus, our results represent an independent discovery, at the 3$\sigma$-level, of the two populations of TWA recently found by \citet{2025A&A...694A..60M}. Moreover, the populations A (UNION-A and L23-A) and B (UNION-B and L23-B) mostly correspond to groups a and b found by \citet{2025A&A...694A..60M}, although with partial entanglement. For example, out of the 42 members of L23-A, 31 and 11 are in common with groups a and b, and out of the 15 members of L23-B, 9 and 6 are in common with groups a and b, respectively.  

The UNION, LACEwING, and BANYAN lists of members are the most contaminated ones, with contamination rates varying between 25\% and 44\%. On the other hand, the most recent membership lists only have up to 15\% of contaminants, although with varying RV coverages.

The analyses of the majority of the membership lists produce significant, at the $2\sigma$ level, detections of expansion in the three components of the expansion rate. It is only in the membership lists by LACEwING, \cite{2019MNRAS.486.3434L}, and L23-B that the expansion is significantly detected in two components, and only in the membership list of UNION-B and \citet{2025A&A...694A..60M} for group b that expansion is not significantly detected in any component. Due to the weakness of this expansion signal, the age determination of this group is the most uncertain. Finally, rotation is significantly detected only in $\omega_z$ and in the membership list by \cite{2019MNRAS.486.3434L}.

Concerning the inferred ages, we observe that, except for the results of the pre-\textit{Gaia} membership lists of BANYAN and LACEwING, the modern membership lists of the classical TWA (i.e. excluding the substructures) by \citet{2019MNRAS.486.3434L}, \citet{2023AJ....165..269L}, and \citet{2025A&A...694A..60M}  result in compatible and precise age determinations. 

The ages of the substructures show that they are coeval, with small $\lesssim$3 Myr age differences. On one hand, the age of group A is well constrained by the three membership lists, with the three expansion rate components significantly detected and age uncertainties between 0.8 and 1.2 Myr. On the other hand, the age uncertainty of group B varies between 1.3 and 2.4 Myr, and is only well constrained with the membership list L23-B, which has two components of the expansion rate significantly detected. Given the  large number of members and RV completeness of the membership lists L23-A and L23-B, we choose these lists to report the ages of the two TWA populations. 

\subsection{32 Orionis and 118 Tau}
\label{results:THOR_118TAU}

We analyse the presence of supra and substructures with the following three membership lists. First, the list compiled by \citet{2022AJ....164..151L}, which contains not only the classical 118TAU and THOR associations, but also new members lying within the shared phase-space between these two. Second, the UNIONS-L22 list, that is the unification of THOR's and 118TAU's UNION lists but excludes the one by \citet{2022AJ....164..151L}. Third, the UNIONS+L22 list, which is the addition of the previous two lists. As done with other lists resulting from the union of several lists (see Sect. \ref{data:membership}), we removed the duplicated sources. We notice that the UNIONS-L22 list has six members in common, thus indicating that the literature lists of the classical THOR and 118TAU already showed some level of entanglement in the studies previous to that by \citet{2022AJ....164..151L}. Our substructure identification methods (see Sect. \ref{methods:substructures}) applied to the previous three membership lists produced the following results.

In \citet{2022AJ....164..151L} list of members, the GMM with two components, which we call L22-A and L22-B, was inferred with non-negligible weights of 0.57 and 0.43, respectively. In this mixture, component L22-A fits the phase-space distribution of the classical members from THOR and 118TAU, while L22-B fits the newly identified members by \citet{2022AJ....164..151L}. These two Gaussian components are mutually exclusive, with Mahalanobis distances of $\rm{\mathcal{M}_{AB}=2.4}$ and $\rm{\mathcal{M}_{BA}=6.3}$, which indicates that the null hypothesis of a single population within \citet{2022AJ....164..151L} list of members can be excluded with a confidence level of 98\%. Given the previous results, we inferred a GMM with three components, which resulted in one of the components with a negligible weight and the other two components being equivalent to those of the GMM with 2 components. Therefore, we conclude that, given the list of members by \citet{2022AJ....164..151L} and his original RV data, we cannot reject this author's claim about THOR and 118TAU belonging to a single population. However, our substructure identification methodology allows us to identify the structure L22-B as a new and independent population. 

In the UNIONS-L22 list of members, the GMM with two Gaussian components, which we call A and B, was inferred with non-negligible weights of 0.27 and 0.73, respectively. The Mahalanobis distances between these components take values of $\rm{\mathcal{M}_{AB}=1.67}$ and $\rm{\mathcal{M}_{BA}=16}$ sigma, which do not fulfil our substructure  identification criterium of at least 2$\sigma$ level. Therefore, we stop our exploration and conclude that, given the UNION-L22 list of members and its GLASSS RV data, THOR and 118TAU remain entangled. 

In the UNIONS+L22 list of members, the GMM with two components, which we call A and B, was inferred with weights of 0.63 and 0.37, respectively. The Mahalanobis distances between these components are $\rm{\mathcal{M}_{AB}=6.5}$ and $\rm{\mathcal{M}_{BA}=3.5}$ sigma, thus indicating that the two components are mutually exclusive with a high confidence level of 99.9\%. As in the case of the membership list by \citet{2022AJ....164..151L}, component A fits the phase-space distribution of the two classical THOR and 118TAU members while component B fits the extended population corresponding to L22-B (more below). The inference of the GMM with three components, called A, B, and C, resulted in non-negligible weight for the three of them and mutually exclusive Mahalanobis distances in all cases, with the smallest distance being $\rm{\mathcal{M}_{AC}=2.7}$, equivalent to confidence level of 99.4\%. The inference of a four-component GMM returned a negligible weight for the fourth component. Thus, we reject the hypothesis of four populations within this membership list. 

The previous phase-space exploration showed that the precise \textit{Gaia} DR3 data allowed us to disentangle three independent populations depending on the membership list and RV data. Although THOR and 118TAU remained entangled in the \citet{2022AJ....164..151L} and UNIONS-L22 lists, in the largest and with the most complete RV data of UNIONS+L22, we were able to disentangle them with a confidence level of 99.99\% equivalent to 4.95$\sigma$. Therefore, we decided to report the three populations identified in the UNIONS+L22 list. From these populations, component A shares 33 members (out of 35) with the 57 literature members of 118TAU, component B shares 42 members (out of 45) with the 71 literature members of THOR, and component C shares 54 members (out of 64) with the 62 members of L22-B. Therefore, we identify components A and B with 118TAU and THOR, respectively. Component C is a new independent population that we name  H\'aap\footnote{The members of component C cover the sky region that, in the cosmogony of the Seri people, corresponds to the constellation of the bura deer that they name H\'aap \citep[see Stellarium,][and references therein to the Seri constellations]{Stellarium}.}. 

Concerning the number of members, we observe that both THOR and 118TAU have literature membership lists with only a few dozen members, resulting in poorly constrained age estimates. The contamination reached maximum values of 41\% in THOR's lists by LACEwING and BANYAN, and 30\% in 118TAU's list by \cite{2022ApJ...939...94M}. The RV coverage is >50\% in both associations.

Concerning the kinematic patterns in the classical 118TAU and THOR membership lists, we only detect, at the 2$\sigma$ level, expansion along the Y and Z directions in four lists, and rotation in the X direction in only one list. On the contrary, in the 118TAU+THOR lists, expansion was significantly detected in the three directions from almost all lists, except in those of L22-A and UNIONS+L22-A. Rotation was detected in X, Y, and Z in the lists by \citet{2022AJ....164..151L} and UNIONS+L22. Given the entanglement of substructures in these lists, we interpret these  detections as the result of motions between structures rather than the true rotation of a single population. 

Concerning the expansion ages of the identified populations and substructures, we observe that ages vary greatly with membership lists. For example, in the case of 118TAU, the BANYAN list provides an age estimate of $17.6\pm7.4$ Myr that is inconsistent with the $26.2\pm5.1$ Myr and $30.1\pm4.7$ Myr inferred from the lists by SPYGLASS and \citet{2022ApJ...939...94M}, respectively. We interpret these large variations as resulting from the entanglement of the THOR and 118TAU groups, which was partially present in the classical literature membership lists. Due to the previous reasons, we decided to report the ages of 118TAU and THOR as inferred from the UNIONS+L22-A and UNIONS+L22-B lists, respectively, which are cleaner, larger, and with more RV coverage than those from the literature. In the same way, we report the age of H\'aap as that inferred from the UNIONS+L22-C membership list. 

\subsection{$\beta$-Pictoris}
\label{results:BPIC}

We test the hypothesis of two populations within the UNION, \citet{2019MNRAS.486.3434L}, and \citet{2024AJ....168..159L} membership lists. We extend our search for substructures to the last two lists due to their large number of members (more than a hundred) compared to the rest of the lists, which only have a few dozen. The results of this exploration are the following.

In the membership list by \citet{2019MNRAS.486.3434L}, the GMM with two components was inferred with a negligible weight for the second component. Thus, we conclude that the null hypothesis of a single population cannot be rejected based on this membership list. 

In the membership lists by \citet[][]{2024AJ....168..159L}, the GMMs with two components allow us to reject the null hypothesis of a single population with confidence levels of 99.2\% and 99.5\% for the original and GLASSS spectroscopic radial velocities, respectively. These two populations, which we call Luhman2024-A and Luhman2024-B, have non-negligible weights of 0.65 and 0.35, respectively. We tested the alternative hypothesis of three populations by inferring GMMs with three components on the original and GLASSS spectroscopic radial velocities. Although the weights of these three components were non-negligible, their mutual Mahalanobis distances did not allow us to accept the alternative hypothesis due to confidence levels of only 93.5\% and  76.8\% in the original and GLASSS data, respectively. 

The identified substructures of Luhman2024-A and Luhman2024-B have 118 and 63 members, respectively. We cross-matched these lists with those from the literature, and we observed that all the literature lists share the majority of their members with Luhman2024-A and a minority with Luhman2024-B, with membership lists that share the most members with Luhman2024-B are those of  \citet[][with 32]{2019MNRAS.486.3434L}, BANYAN (with 18) and \citet[][with 13]{2019MNRAS.489.3625C}, with the rest of the lists sharing less than 10 members. Furthermore, we searched the CDS \citep{2000A&AS..143....9W} for possible identification of parent groups for each of the Luhman2024-B members and found that all of them belong to BPIC. Thus, we conclude that Luhman2024-B is a new population, which was partially entangled with the members of BPIC, and that we call Balaam due to the fact that its brightest member, HD191089, lays at the head of the Mayan constellation thus called \citep[see Stellarium,][ and  references to the Maya constellations]{Stellarium}. 

Concerning Luhman2024-A, we observe that it contains the majority of the classical BPIC members from the literature, although the membership list by \citet{2024AJ....168..159L} greatly extended its census. Our hierarchical tree exploration for phase-space substructures showed that this population can be modelled with a GMM of two components with non-negligible weights of 0.73 and 0.27 for components Luhman2024-AA and Luhman2024-AB, respectively, although with Mahalanobis distances that do not allow us to reject the null hypothesis of a single population due to a confidence level of only 82\%. Due to convergence issues in the age determination of Luhman2024-A (more below), instead, we provide the age estimates of its substructures Luhman2024-AA and Luhman2024-AB.

Concerning the UNION list, we tested the alternative hypothesis of two populations and rejected it due to a Mahalanobis distance $\rm{\mathcal{M}_{BA}=1.9}$, which fails to fulfil our $2\sigma$ criterion. Similarly, when testing the alternative hypothesis of three populations, the third component remained entangled with the other two, with Mahalanobis distances smaller than 2.

When applying our expansion age methodology to the lists of members by the literature and identified phase-space structures, we observed that the sampling algorithm failed to converge on the membership lists by \citet{2024AJ....168..159L} with the original RV, the UNION one, and the Luhman2024-A. These convergence failures have their origin in several factors. First, BPIC is the closest stellar association to the Sun, thus the astrometric uncertainties of its members are extremely low and the simple linear velocity may not be flexible enough to describe the phase-space complexity of this group \citep[see Sect. 3.3 of][]{2025A&A...693A..12O}. Second, the presence of substructures that, although they fail to pass our strict criteria to identify them as populations, nonetheless have complex phase-space characteristics that prevent their modelling. For the previous reasons, in the following, we do not present the properties of those lists where our expansion age methodology failed.

With respect to the number of members, we observe that the most contaminated membership lists are those of LACEwING and \citet{2019MNRAS.486.3434L} with 45\% and 17\% of contaminants, respectively, while the least contaminated ones are those of \cite{2020A&A...642A.179M}, \cite{2019MNRAS.489.3625C}, \cite{2023ApJ...946....6C}, and \cite{2022ApJ...939...94M}. The RV coverage varies from 52\% to 95\%, indicating that future increases of the RV coverage could improve the age determination of this association.

Concerning the detection of statistically significant kinematic patterns, we observe that contraction along the Z direction is observed in the membership lists by SPYGLASS, \citet{2020A&A...642A.179M} with the GLASSS data, and Luhman2024-AB. On the contrary, the majority of the membership lists result in significant detections of expansion along the X and Y directions, with the only exceptions being the SPYGLASS list where no significant expansion was detected, and those of LACEwING, \cite{2022ApJ...939...94M}, and Luhman2024-AA, where expansion is only detected along the X direction. With respect to rotation, it is detected in seven membership lists along the X direction, in two along the Y direction, and only in one along the Z direction. 

Concerning the expansion ages, all our age estimates are compatible (within the 95\% HDI) and consistent within 1$\sigma$, except for the case of Balaam (Luhman2024-B), which has the youngest age estimate, and is only consistent, at 1$\sigma$, with the ages resulting from the lists by \citet{2019MNRAS.489.3625C} and \citet{2020A&A...642A.179M}. The rest of the age estimates vary from 23.4 Myr to 28.5 Myr. We notice that the age posterior distribution resulting from the membership lists by BANYAN, \citet{2019MNRAS.486.3434L}, \cite{2019MNRAS.489.3625C},  \citet{2024AJ....168..159L}, and Luhman2024-AB are all bimodal, with a first mode peaking at ages younger than 20 Myr and the second one peaking at ages older than 25 Myr. Not surprisingly, the membership lists by \citet{2019MNRAS.486.3434L}, BANYAN and \cite{2019MNRAS.489.3625C} are the ones sharing more members with the Balaam (Luhman2024-B) population. Thus, we interpret the youngest peak as arsing from partial entanglement with the Balaam population. Moreover, the young secondary age peak of substructure Luhman2024-AB at 16 Myr indicates that this group may still be partially entangled with Balaam (Luhman2024-B). In the future, complete RV coverage will help to not only improve the age determination of these populations but also to  fully disentangle them. 

As the expansion age of BPIC, we decided to report the value inferred from the membership list by \citet[][with their original RV]{2020A&A...642A.179M} because it is the one resulting in the smallest uncertainty, of 4.8 Myr. Although the membership list by \citet{2024AJ....168..159L} returns an uncertainty of 4.7 Myr, this value results from a list containing the substructures Luhman2024-A and Balaam (Luhman2024-B). Furthermore, we also report the expansion age of the newly identified Balaam population.

\subsection{Octans}
\label{results:OCT}

Our search of substructures in the UNION list of members, which is the most populated one, resulted in no rejection of the null hypothesis of a single population. The GMM fitted with two components returned no-negligible weights of 0.51 and 0.49 for components A and B, respectively, and Mahalanobis distances of $\rm{\mathcal{M}_{AB}=4.3}$ and $\rm{\mathcal{M}_{BA}=1.9}$. Although, according to our criteria (see Sect. \ref{methods:substructures}), these distances are not sufficient to reject the null hypothesis, the 1.9$\sigma$ value is at the limit of our criterion. When we attempted to estimate the expansion age of the UNION list as a single population, the sampling algorithm faced convergence issues. The same occurred when we attempted to infer the age of the UNION-B substructure. Thus, we decided to test the hypothesis of three populations by fitting a GMM with three components. It resulted in non-negligible weights of 0.39, 0.50, and 0.11 for the components that we call UNION-A, UNION-B, and UNION-C, respectively. The mutual Mahalanobis distances between components A and B indicate that these two groups are partially entangled ($\rm{\mathcal{M}_{AB}=1.6}$ and $\rm{\mathcal{M}_{BA}=2.7}$). On the contrary, component C was identified as an independent population with confidence levels larger than 99.5\% (i.e. Mahalanobis distances of $\rm{\mathcal{M}_{AC}=19.8}$, $\rm{\mathcal{M}_{CA}=15.7}$, $\rm{\mathcal{M}_{BC}=14.3}$ and $\rm{\mathcal{M}_{CB}=2.9}$). Cross-matching these groups with the census of stellar systems done by \cite{2023A&A...673A.114H}, we found that out of  the 69, 76, and 17 members of UNION-A, UNION-B and UNION-C, respectively, seven of UNION-A correspond to HSC 1998, six of UNION-B correspond to groups HSC 1998 and CWNU 1178, and nine of UNION-C to HSC 2597. Thus, the independent identification of these groups confirms the partial entanglement of substructures UNION-A and UNION-B and the independent population of UNION-C, for which we refer as HSC2597. In the following, we report the expansion ages and properties of the OCT substructures UNION-A and UNION-B, and HSC 2597 (UNION-C).

Our age methodology failed to provide an age estimate for the substructure UNION-B due to convergence issues. After performing the recommended steps in \citet{2025A&A...693A..12O}, we were unable to arrive at convergence of the age models. For this reason, we do not include this list in the following analysis. 

Concerning the properties of the membership lists, we observe that the pre-\textit{Gaia} ones have a large number of contaminants (between 43-45\%) and low RV coverages (45-66\%). On the contrary, the lists by \citet{2024A&A...689A..11G} and \cite{2022ApJ...939...94M} have low number of contaminants ($\sim$6\%) and large RV coverages (87-100\%). 

Concerning the kinematic patterns, all the membership lists, except that of HSC2597 (UNION-C), present significant detections of expansion along the X and Y components. In addition, the lists by BANYAN, \citet{2022ApJ...939...94M}, and UNION-A present significant contraction along the Z direction. Rotation is significantly detected in the Y and Z components, in the LACEwING and \citet{2022ApJ...939...94M}, and only in Z in the UNION-A.

Concerning the expansion ages, we observe that all OCT membership lists result in consistent and compatible age estimates that vary between 29-34 Myr, with uncertainties in the 4-6 Myr interval. The older age of HSC 2597 (UNION-C) supports the hypothesis of it as an independent population. 

We decided to report the expansion age inferred from the UNION-A list, given its large number of members and RV coverage. In the case of the independent population of HSC2597 we report its age based on the UNION-C list.

\subsection{Columba}
\label{results:COL}

We search for phase-space structures in the UNION and BANYAN membership lists. We include the BANYAN one due to convergence issues in the sampling algorithm of our methodology (thus it is not included in the analysis as an independent list). The search for substructures resulted in the identification of populations BANYAN-A and BANYAN-B with non-negligible weights of 0.82 and 0.18, respectively. These populations were identified with confidence levels greater than 99.95\% (equivalent to Mahalanobis distances larger than 3.5). Our search of substructures in the UNION list resulted in similar conclusions as those obtained in the BANYAN one. We also reject the null hypothesis of a single population with a confidence level of 99.99\%, but this time, the weights of components UNION-A and UNION-B are 0.89 and 0.11, respectively. We tested the alternative hypothesis of three populations by fitting a GMM with three components. The three components A, B, and C have 126, 46, and 15 members, respectively. Nonetheless, the three populations hypothesis was rejected in favour of the two populations due to a partial entanglement between components A and B, which have Mahalanobis distances of $\rm{\mathcal{M}_{AB}=1.5}$ and $\rm{\mathcal{M}_{BA}=7.4}$. The 15 members of UNION-C have 13 members in common with BANYAN-B, and only two and one counterparts in the LACEwING and \citet{2019MNRAS.486.3434L} membership lists. Cross-matching with the large and complete survey by \cite{2023A&A...673A.114H}, we only identify one crossmatch with system HSC 1347. It does not corresponds to BANYAN-B or UNION-C due to its discrepant distance, which the previous authors report as 85 pc, whereas UNION-C is located at only 31 pc. Thus, the discrepant distance and lack of counterparts in the membership lists from the \textit{Gaia} era allow us to conclude that BANYAN-B (UNION-C) is a new population related to but independent of COL. Following the nomenclature of its brightest member, we call this new population the $\omega$ Aurigae association, and abbreviate it as OMAU. The membership lists of the substructures UNION-A and UNION-B have 16 and 32 members in common with the catalogue by \cite{2023A&A...673A.114H}. On one hand, UNION-A has seven, four, three, and two members in common with HSC 1900, HSC 1766, Alessi 13, and $\beta$-Tucanae, respectively. On the other hand, UNION-B has 27 and five members in common with HSC 1900 and HSC 1684. These cross-matches prove that COL has a complex phase space.

We observe that the membership lists of BANYAN, LACEwING and \cite{2019MNRAS.486.3434L} are the most contaminated (14-36\%) ones, while those of \cite{2022ApJ...939...94M} and \cite{2024AJ....168..159L} have no contaminants. The RV coverage varies from 50-100\%.

Concerning the kinematic patterns of COL, there is significant expansion along the X or Y components in all lists and contraction in the Z component from the membership lists by \citet{2019MNRAS.486.3434L}, \citet{2024AJ....168..159L}, BANYAN-A, UNION-A and UNION-B. Rotation is detected in the X, Y, and Z components on the membership lists of LACEwING, \citet{2019MNRAS.486.3434L}, and BANYAN-A respectively, as well as in UNION-A and UNION-B, in Z and X and Y, respectively. In OMAU, there are no significant expansions or contractions but only rotation in Y from the BANYAN-B list.

Concerning the expansion ages, we observe that all values are compatible (within the 95\% HDI) and consistent (within 1$\sigma$), except the two values from the OMAU population, which are older than COL. We notice that the posterior age distributions of COL's substructures are bimodal, with one peak at 21 Myr and another at 39 Myr, which will be further discussed in Sec. \ref{discussion:COL}. We decided to report the expansion age of COL as inferred from the membership list UNION-A because of its large number of members and significant expansion in X and Y directions. In the case of OMAU, we report its age from the UNION-C list because it has a smaller uncertainty.

\subsection{Carina}
\label{results:CAR}

We search for phase-space substructures in the memberships by \citet[][with both the original and GLASSS spectroscopic radial velocities]{2024AJ....168..159L} and the UNION one. In the UNION list, the null hypothesis cannot be rejected due to a negligible weight for the second component. However, the results on the membership list by \citet{2024AJ....168..159L} of the CAR-Ext population show that the null hypothesis of a single population can be rejected with high confidence levels of 99.99\% in both GLASSS and original RVs. The weights of components Luhman2024-A and Luhman2024-B are 0.79 and 0.21 in both GLASSS and original datasets. Similarly, their Mahalanobis distances are $\rm{\mathcal{M}_{AB}=7.5}$ and $\rm{\mathcal{M}_{BA}=6.1}$ in the GLASSS data and $\rm{\mathcal{M}_{AB}=7.4}$ and $\rm{\mathcal{M}_{BA}=6.5}$, in the original data. Due to the extremely large population of Car-Ext and the computational burden of a three and four-component GMM, instead of testing these alternative hypotheses, we proceed hierarchically on the already discovered populations. This hierarchical exploration shows that both populations have complex phase-space distributions. On one hand, population Luhman2024-A has two substructures, Luhman2024-AA and Luhman2024-AB that, although have non-negligible weights of 0.56 and 0.44, respectively, their Mahalanobis distances of $\rm{\mathcal{M}_{AB}=7.3}$ and $\rm{\mathcal{M}_{BA}=1.9}$ do not fulfil our criteria of 2$\sigma$. On the other hand, population B has two components, Luhman2024-BA and Luhman2024-BB, with non negligible weights of 0.19 and 0.81 and Mahalanobis distances of $\rm{\mathcal{M}_{AB}=14.7}$ and $\rm{\mathcal{M}_{BA}=19.9}$ that allow us to conclude, with confidence levels larger than 99.99\%, that these components are populations rather than substructures.

We searched the literature for counterparts to the previous populations and substructures. On one hand, the members of substructures Luhman2024-AA and Luhman2024-AB are partially entangled with the rest of CAR's literature members and a cross-match with the catalogue by \citet{2023A&A...673A.114H} shows that Luhman2024-AA has 200 members in common with Platais 8 and Luhman2024-AB has three members in common with Platais 8 and another three with HSC 2399. On the other hand, the members of the populations Luhman2024-BA and Luhman2024-BB appear unrelated to CAR's literature lists, with only one member of Luhman2024-BB in common with the membership list by \citet{2019MNRAS.486.3434L}. According to the CDS, two members of Luhman2024-BA pertain to IC2395, and one, one, and three members of Luhman2024-BB pertain to NGC 2669, Vela R1 association, and Platais 8, respectively. The cross-match with the catalogue by \citet{2023A&A...673A.114H} shows that Luhman2024-BA and Luhman2024-BB have 20 and 30 members in common with group HSC 2139, which the previous authors cite as having only 67 members. Thus, we conclude that the substructure Luhman2024-AA and Luhman2024-AB corresponds to the open cluster Platais 8 and the classical CAR association, respectively. On the other hand, the populations Luhman2024-BA and Luhman2024-BB, which were previously related to CAR and HSC 2139, are now two independent associations. Moreover, from the 129 final members in the UNION list, 12 are in common with Luhman2024-AA, 25 with Luhman2024-AB, zero with Luhman2024-BA, and one with Luhman2024-BB, thus confirming our classification of Luhman2024-AB as the classical CAR association, its entanglement with Platais 8, and the independence of Luhman2024-BA and Luhman2024-BB. The members of Luhman2024-BA and Luhman2024-BB are located in the modern constellations of Vela and Carina, respectively. This sky region is covered by a Maya constellation \citep[see Stellarium,][ and references therein to the Mayan constellations]{Stellarium} depicting Nal, the corn's god, travelling in a canoe, which is called chem in the Maya language. Due to this correspondence, we decided to call Luhman2024-BA and Luhman2024-BB as the Nal and Chem associations, respectively.

Concerning the number of members and their properties, we observe that the literature membership showed low levels of contaminants, varying from 13\% in LACEwING to 5\% in BANYAN. Unfortunately, the RV coverage remains at 65\% or less. In the case of the list by \citet{2024AJ....168..159L}, which was treated independently, it has low contamination (4\%) but also a low RV coverage of 40\%.

Concerning the kinematic patterns, the majority of the membership lists present contraction in the Z direction, except for those of \citet{2024AJ....168..159L}, Luhman2024-B, and Luhman2024-BB. Expansion is significantly detected, in either the X or Y direction, in all lists except in Luhman2024-BA. Rotation is detected in the X direction in the lists by BANYAN, \cite{2019MNRAS.486.3434L}, UNION, \citet{2024AJ....168..159L}, and Luhman2024-AB, in the Z direction in the lists by UNION, Platais 8 (Luhman2024-AA), and Luhman2024-AB, and in the Y direction in \citet{2024AJ....168..159L}.

The expansion ages of CAR (excluding the Platais 8, Nal and Chem populations) are all consistent within the uncertainties. We decided to report the age inferred from the Luhman2024-AB list due to its large number of members and significant detections of expansion along the X and Y directions. We also report the expansion ages of Platais 8, Nal, and Chem, from the lists by Luhman2024-AA, Luhman2024-BA, and Luhman2024-BB, respectively.

\subsection{Tucana-Horologium}
\label{results:THA}

In THA, our methodology failed to converge in several membership lists, except in those by BANYAN, \citet{2019MNRAS.486.3434L}, and \citet{2023MNRAS.520.6245G}. This failure may have its origin in the close proximity (20 pc) and large extent (about 60 pc) of its members, in combination with the lack of flexibility of the linear velocity field model \citep[see Sect. 3.3 of][]{2025A&A...693A..12O}. Thus, in the following, we present the results of the membership lists in which convergence of the sampling algorithm was ensured, and we leave to future work the analysis of the large membership lists by \citet{2024AJ....168..159L} and the other authors.

We searched for phase-space substructures and populations in the three membership lists where our methods did not fail. In the list by \citet{2023MNRAS.520.6245G}, with both the original and GLASSS RV data, the GMM with two components were inferred with negligible weights for the second component, thus indicating that the alternative hypothesis of two substructures was rejected. In the BANYAN list, the GMM with two components returned non-negligible weights 0.79 and 0.21 for components BANYAN-A and BANYAN-B, respectively, with Mahalanobis distances of $\rm{\mathcal{M}_{AB}=6.2}$ and $\rm{\mathcal{M}_{BA}=1.2}$, thus, we cannot reject the null hypothesis of a single population but we provide ages for the two substructures as well. Similarly, in the list by \citet{2019MNRAS.486.3434L}, the GMM with two components returned non-negligible weights of 0.77 and 0.23 for components Lee+2019-A and Lee+2019-B, respectively, with Mahalanobis distances of $\rm{\mathcal{M}_{AB}=2.4}$ and $\rm{\mathcal{M}_{BA}=0.4}$, thus we provide the ages of these two substructures. The substructures BANYAN-A and Lee+2019-A share 18 and 16 members with the list by \citet{2023MNRAS.520.6245G}, respectively, while substructures BANYAN-B and Lee+2019-B share only three and one members, respectively. We searched at the CDS and found that BANYAN-A has 66 of its 70 members as pertaining to THA and the following systems: BPIC (1), AB Dor (1), Alessi 13 (2), COL (4), Pleiades moving group (5), and Tucana moving group (6). Similarly, BANYAN-B has its 15 members pertaining to THA and to the following systems: CAR (1), Alessi 13 (2) and Tucana moving group (3). In Lee+2019-A, 159 out of its 165 members are cited in the CDS as members of THA and to the following systems: Tucana moving group (84), Pleiades moving group (6), Alessi 13 (2), COL (2), AB Dor (1), and BPIC (1). Finally, BANYAN-A, BANYAN-B, Lee+2019-A and Lee+2019-B have all of its cross-match with the catalogue by \citet[][47, 10, 146, and 32, respectively ]{2023A&A...673A.114H} as belonging to the $\beta$-Tucanea group (i.e. THA). Thus, we conclude that given the minimum entanglement of these substructures with other literature groups and the lack of evidence supporting the hypothesis of two populations, we consider them as substructures of THA. 

Concerning the contaminants and RV coverage of the three membership lists, we observe that the one by \citet{2019MNRAS.486.3434L} is the most contaminated, with 10\%, while the other two are virtually clean. The RV coverage varies from 95\% in \citet{2023MNRAS.520.6245G}, 63\% in BANYAN, to 56\% in \citet{2019MNRAS.486.3434L}.

Concerning the kinematic patterns, we observe that significant contraction is observed along the Z direction only and in the lists of group A (i.e BANYAN-A and Lee+2019-A). On the other hand, significant expansion is observed only in the X direction from the lists by \cite{2023MNRAS.520.6245G} with the original RV, BANYAN-A, Lee+2019-A, and Lee+2019-B. No significant rotation was detected in any of the membership lists.

With respect to the expansion ages, we observe that the values inferred from all lists are consistent within the uncertainties, except for those from \citet{2019MNRAS.486.3434L} and Lee+2019-B, which are the older ones. However, Lee+2019-A is younger than the later group and consistent with the age estimates from the other lists, which indicates that Lee+2019-B is most likely part of an older group that our method and data do not allow to disentangle from group Lee+2019-A.

We decided to report the expansion ages from the substructures Lee+2019-A and Lee+2019-B because these are the ones with the largest sample of members and RV, and have significant expansion along the X direction and the smallest uncertainties.  

\section{Discussion}
\label{discussion}

Once our results were presented in the previous section, in this one we discuss them. As before, we start with a brief overview of the most general results and their implications. Then, we do  specific discussions for each LYSA. Finally, we discuss the caveats of our methods and their implications in our results.

\subsection{Overview}
\label{discussion:overview}

Our findings can be categorised into the following: the expansion ages themselves, the populations and substructures present in the literature membership lists of the classical associations, and the significant detections of kinematic patterns of contraction, expansion, and rotation. We now briefly discuss these three categories.

The ages that we report here represent the largest and most homogenous compilation of LYSAs' ages obtained through the expansion rate method. This compilation will serve not only to study the associations themselves but a wide interval of phenomena, varying from the characterisation of exoplanets and disks to the star formation history in the solar neighbourhood, including understanding the differences between the various dating techniques, as done, for example, in  \citet{2024NatAs...8..216M}.

The populations and substructures recovered or discovered within the literature membership lists of the classical associations serve as a probe that the stellar systems in the solar neighbourhood possess much phase-space complexity that was thought before. The unveiling of this complexity was possible not only due to the exquisite precision of the \textit{Gaia} data but also to the development in recent years of machine learning methodologies working beyond the traditional observational spaces \citep{2023A&A...677A..59R,2025A&A...693A..12O}. The identification of populations or substructures not only serves to improve our understanding of the solar neighbourhood but also to obtain more exact estimates of their parameters, although in some cases implies a degradation of precision due to low-number statistics. In this sense, future work will be needed to increase their census of members or the precision of their observational data.

Finally, our compilation of kinematic patterns for the 18 LYSAs is the largest and most homogeneous one inferred in the Galactic phase-space reference system. Other works have presented this type of compilation focused on open clusters \citep[e.g.][]{2023A&A...673A.128G,2024A&A...687A..89J} and independently obtained from observational subspaces of proper motions and RV. In Olivares et al. (in prep.) we explore the correlations between the various kinematics patterns detected here with age and Galactic position.

\subsection{$\epsilon-$ and $\eta-$ Chamaeleontis}
\label{discussion:EPCHA_and_ETCHA}

\begin{figure}[ht!]
\centering
\includegraphics[width=\columnwidth]{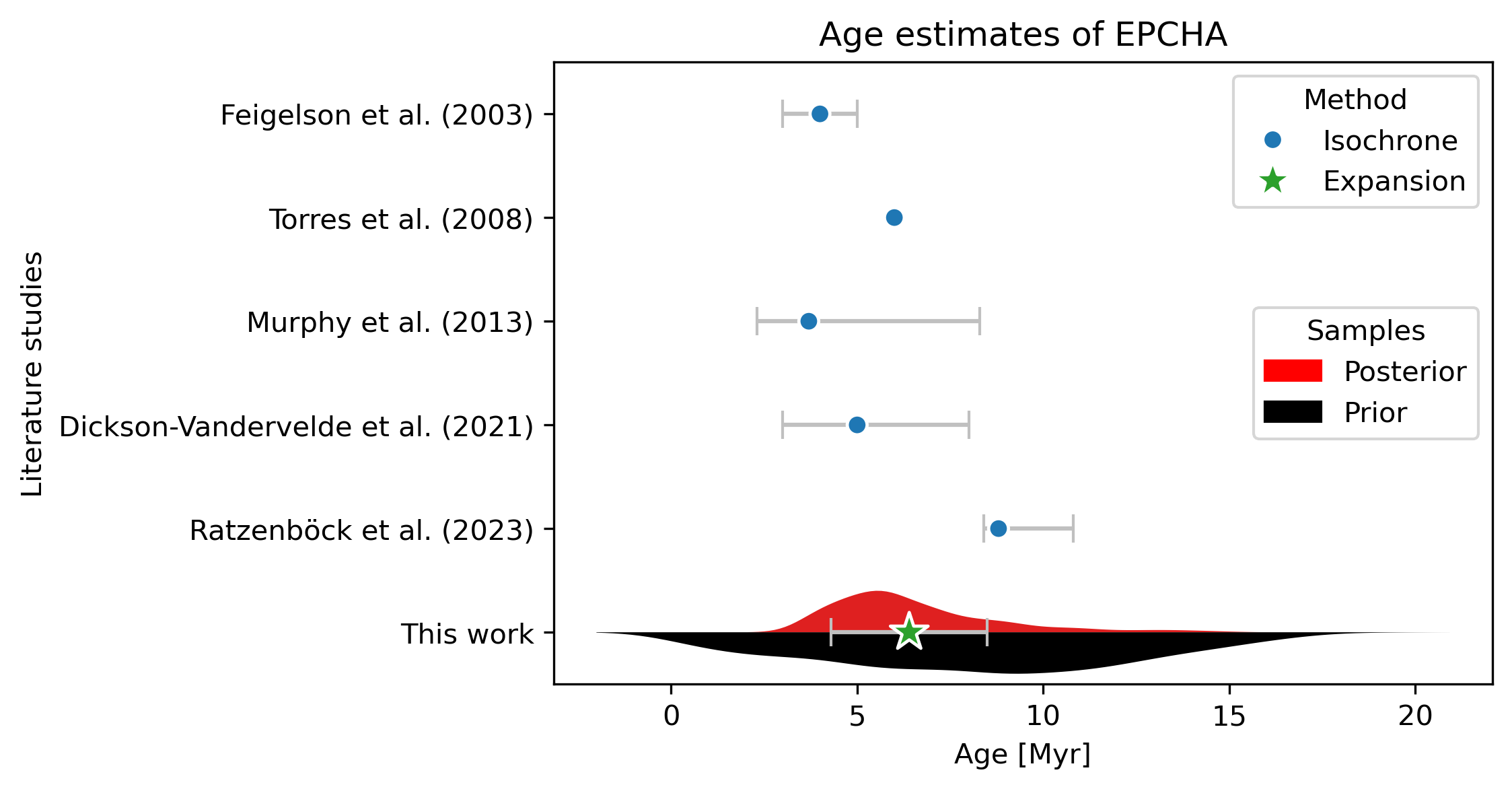}
\caption{Age distribution for EPCHA. The violin plot shows kernel density estimates obtained from samples of the prior (bottom) and posterior (top) distributions. Previous literature age estimates are included for comparison.}
\label{figure:EPCHA_lit}
\end{figure}

\begin{figure}[ht!]
\centering
\includegraphics[width=\columnwidth]{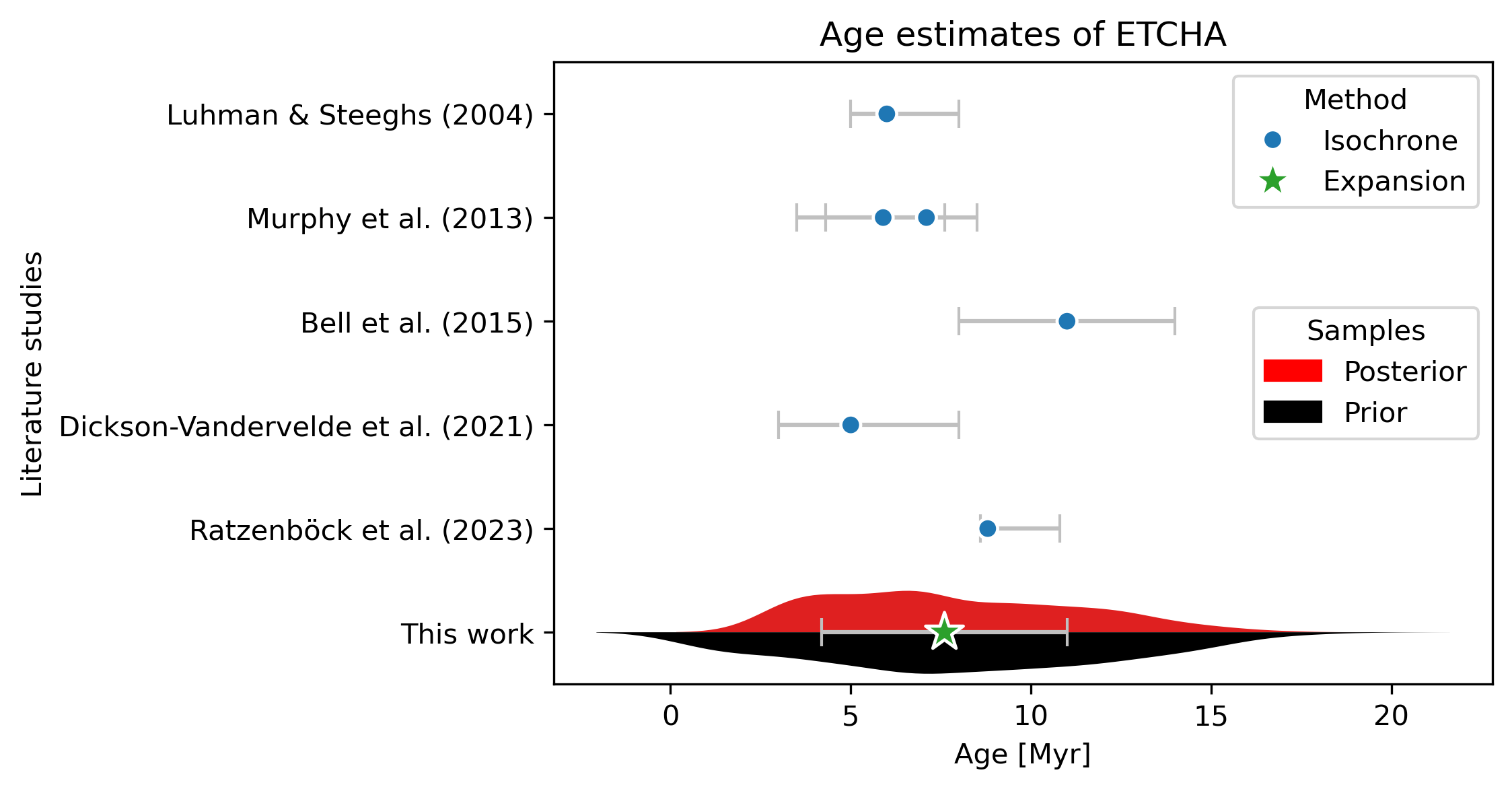}
\caption{Age distribution for ETCHA. The violin plot shows kernel density estimates obtained from samples of the prior (bottom) and posterior (top) distributions. Previous literature age estimates are included for comparison. The two age estimates of \citet{2013MNRAS.435.1325M} correspond to different isochrone models.}
\label{figure:ETCHA_lit}
\end{figure}

EPCHA and ETCHA have been the subject of various literature age estimates, most of them obtained through isochrone fitting. Figures \ref{figure:EPCHA_lit} and \ref{figure:ETCHA_lit} show, colour-coded with the method through which they were obtained, the literature age estimates of EPCHA and ETCHA, respectively. From these figures, we observe the following.

First, due to the low number of members and the scarcity of RV measurements, the age posterior distribution of ETCHA has only mildly shrank with respect to the prior distribution. In the case of EPCHA, the posterior is clearly narrower than the prior distribution. Indeed, the dispersion of EPCHA is 2.1 Myr, and that of ETCHA is 3.4 Myr, which are smaller than that of the prior, 5 Myr. Although in ETCHA this level of shrinking is marginal, it nonetheless suffices to distinguish the posterior from the prior, and thus, to provide an age estimate.

Second, the age posterior distributions of EPCHA and ETCHA have different shapes. While the former has one clear peak, the latter has a long positive tail and hints of three peaks. The asymmetry and irregular shape of ETCHA's posterior distribution suggest that possible phase-space substructures could still be entangled within the UNION list of members. An extended list of members and RV measurements will be needed to further explore this possibility. 

Third, EPCHA's expansion age is compatible (within the 95\% HDI) with previous isochrone age estimates from the literature, and consistent, at the 1$\sigma$ level, with the majority of them, except with those by \cite{2013MNRAS.435.1325M} and \citet{2023A&A...678A..71R}. Even though we are using the same list of members as the later authors, our expansion age ($6.4\pm2.1$ Myr) is inconsistent with their isochrone age ($8.8_{-0.8}^{+2.0}$ Myr), thus indicating that systematic offsets may be present between these two age dating techniques at this young age. 

Fourth, ETCHA's expansion age is consistent, at the 1$\sigma$ level, with all the literature ages. This overall agreement is explained by the posterior similarity with the prior due to the scarcity of members and RV measurements. Future work should focus on extending ETCHA's membership list and improving its RV coverage.

Our phase-space analysis found substructures entangled within the literature lists of members, particularly in EPCHA, where we reidentified the MuscaFG population found by \citet{2023A&A...677A..59R}. Without the disentanglement of these substructures, EPCHA's age determination would have resulted in an older and apparently more precise age of $8.5\pm1.4$ Myr. The previous result is yet another example of the impact that phase-space complexity has on kinematic age determinations. However, we notice that our UNION-B list of members for EPCHA lacks the necessary constraining information to provide a robust age determination, and for this reason, we choose to report the age estimate obtained from the most populous SigMA list.

The coevality of EPCHA and ETCHA has long been hypothesised due to their proximity and similar ages \citep[see, for example, the discussion in Sect. 7.1 of][]{2013MNRAS.435.1325M}. The similar ages and space velocities that we infer here for these two associations (see Table \ref{table:summary} and those in Appendix \ref{appendix:tables}) fully support not only the hypothesis of coevality but also that of joint formation. Recently, \citet{2024RAA....24e5004B} use \textit{Gaia} DR3 data to traceback the trajectories of these two associations, together with those of the Chamaeleon I and II regions and found that the former two intersected $\sim$8 Myr ago and that the four of them were together at the same Galactic height 10-15 Myr ago. Although tracing back to 10-15 Myr is probably to old for these groups, those results also support the hypothesis of EPCHA and ETCHA forming together \citep{2008hsf2.book..757T,2005ApJ...619..945J} out of the same parent molecular cloud as the Chamaeleon star-forming region \citep{2021A&A...646A..46G}. Furthermore, as proposed by \cite{2023A&A...678A..71R}, these fourth groups belong to the Scorpius-Centaurus complex and formed in the third and fourth star formation bursts of this complex, with EPCHA and ETCHA forming 10 Myr ago and Chamaeleon I and II $\sim$5 Myr ago. However, the traceback analysis of \citet{2024RAA....24e5004B} suggests that Chamaeleon I and II are probably much younger, with ages <1 Myr. All the previous hypotheses and results call for a reassessment of the Chamaeleon I and II ages based on larger samples of members and more precise and complete RV data. 

\subsection{Musca-Foreground}
\label{discussion:muscafg}
\begin{figure}[ht!]
\centering
\includegraphics[width=\columnwidth]{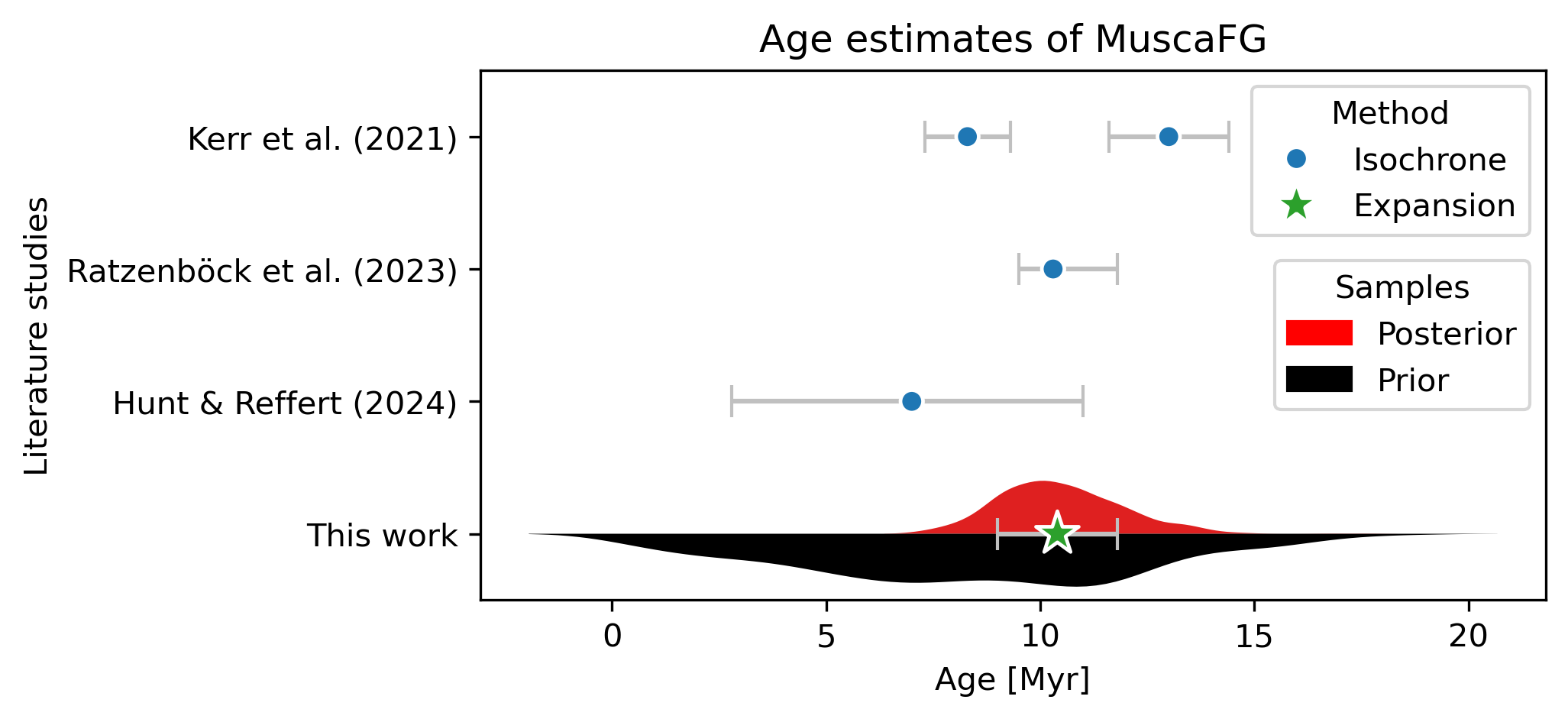}
\caption{Age distribution for MuscaFG. The violin plot shows kernel density estimates obtained from samples of the prior (bottom) and posterior (top) distributions. Previous literature age estimates are included for comparison. The two age estimates by \cite{2021ApJ...917...23K} correspond to those of its SC-27A and SC-27B groups.}
\label{figure:MuscaFG_lit}
\end{figure}

The literature age determinations of MuscaFG are shown in Fig. \ref{figure:MuscaFG_lit}. As can be observed, these estimates are very recent and have only been determined through isochrone fitting. In the cases of \cite{2021ApJ...917...23K} and \cite{2024A&A...686A..42H}, we report two ages corresponding to the substructures in which our lists of members have been identified. In the case of \cite{2021ApJ...917...23K}, our membership list has 26 members in common with their Sco-Cen list, out of which 22 pertain to Lower-Centaurus-Crux and out of these five and four belong to their A (EPCHA) and B groups, respectively, with the remaining one still unclassified. The isochrone ages that these authors report for their groups A and B are $8.3\pm1.0$ and $13\pm1.4$ Myr, respectively. In the case of \cite{2024A&A...686A..42H}, our membership list has four and six members in common with their HSC2523 and HSC2515 groups for which those authors report isochrone ages of $7_{-4.2}^{+4.0}$ and $116_{-100}^{+462}$ Myr, respectively. Due to its extreme value and uncertainty, this latter age is not shown in Fig. \ref{figure:MuscaFG_lit}. Finally, the isochrone age determination by \cite{2023A&A...678A..71R} corresponds to that of their MuscaFG group with 95 members, from which 36 of our 37 MuscaFG members are in common. 

Figure \ref{figure:MuscaFG_lit} shows that our expansion age estimate is (1$\sigma$) compatible with the two latest isochrone age determinations from the literature and falls within the age range of the groups SCA-27A and SCA-27B by \cite{2021ApJ...917...23K}. The differences between our age estimate and those by \cite{2021ApJ...917...23K} and \cite{2024A&A...686A..42H} most likely arise from the different membership lists. While the previous authors use the generic HDBSCAN clustering algorithm to identify members, our method and that of \citet{2023A&A...678A..71R,2023A&A...677A..59R} use clustering algorithms specifically conceived for the astrophysical problem of identifying members and substructures. As a consequence, our age estimate is in excellent agreement with that by \cite{2023A&A...678A..71R} despite our wide and weakly informative prior. 

To the best of our knowledge, this is the first time that an expansion age is reported for the MuscaFG group. This independent age determination supports the star-formation scenario proposed by \cite{2023A&A...678A..71R} in which MuscaFG formed 10 Myr ago as part of the third star formation peak of the Sco-Cen complex.

\subsection{TW Hydrae}
\label{discussion:TWA}
\begin{figure}[ht!]
\centering
\includegraphics[width=\columnwidth]{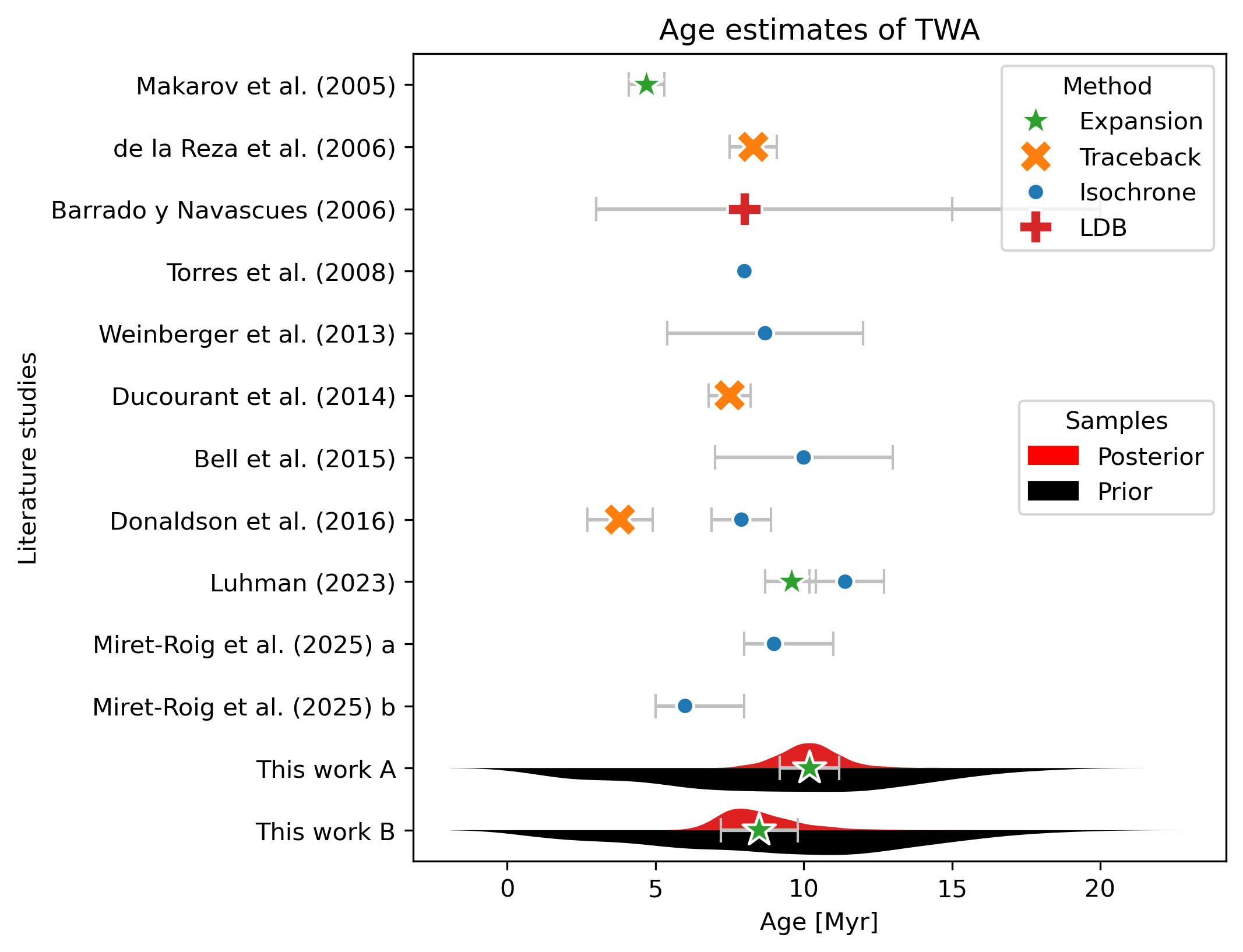}
\caption{Age distribution for TWA A and B. The violin plots show kernel density estimates obtained from samples of the prior (bottom) and posterior (top) distributions. Previous literature age estimates are included for comparison.}
\label{figure:TWA_lit}
\end{figure}

The age of TWA has been the subject of several literature studies that have obtained it through dynamical and nuclear dating methods. Figure \ref{figure:TWA_lit} provides a review of the previous TWA's age estimates colour-coded by the type of dating method. As can be observed, most of the literature estimates are compatible with an age of 7-10 Myr, except for the dynamical methods of \citet[][ $4.7\pm0.6$ Myr]{2005MNRAS.362.1109M} and \citet[][$3.8\pm1.1$ Myr]{2016ApJ...833...95D}. Our Bayesian expansion age is older than all pre-\textit{Gaia} dynamical age estimates but consistent (within 1$\sigma$) with all the isochrone ones \citep[except with the age of group b, $6_{-1}^{+2}$ Myr by][]{2025A&A...694A..60M}, and with the recent expansion age determination by \citet[][$9.6_{-0.8}^{+0.9}$ Myr]{2023AJ....165..269L}. 

The expansion ages inferred here for substructures A and B (called in Sect. \ref{results:TWA}, L23-A and L23-B, respectively) show a similar age gradient as that reported by \cite{2025A&A...694A..60M} for their populations TWA-a and TWA-b, although with shifts of 1-3 Myr in the absolute age determinations. These shifts cannot be entirely attributed to the different age dating methods, given that the membership lists, although similar, are not identical. 

Summarising, our Bayesian methodology allows us to independently confirm the discovery of the two populations reported by \cite{2025A&A...694A..60M}, their young and similar ages, and their age gradient. Future RV surveys will be fundamental to increase the current 50\% RV coverage of these populations, and thus, further constrain their age uncertainty and gradient.

\subsection{32 Orionis and 118 Tau}
\label{discussion:THOR+118TAU}

\begin{figure}[ht!]
\centering
\includegraphics[width=\columnwidth]{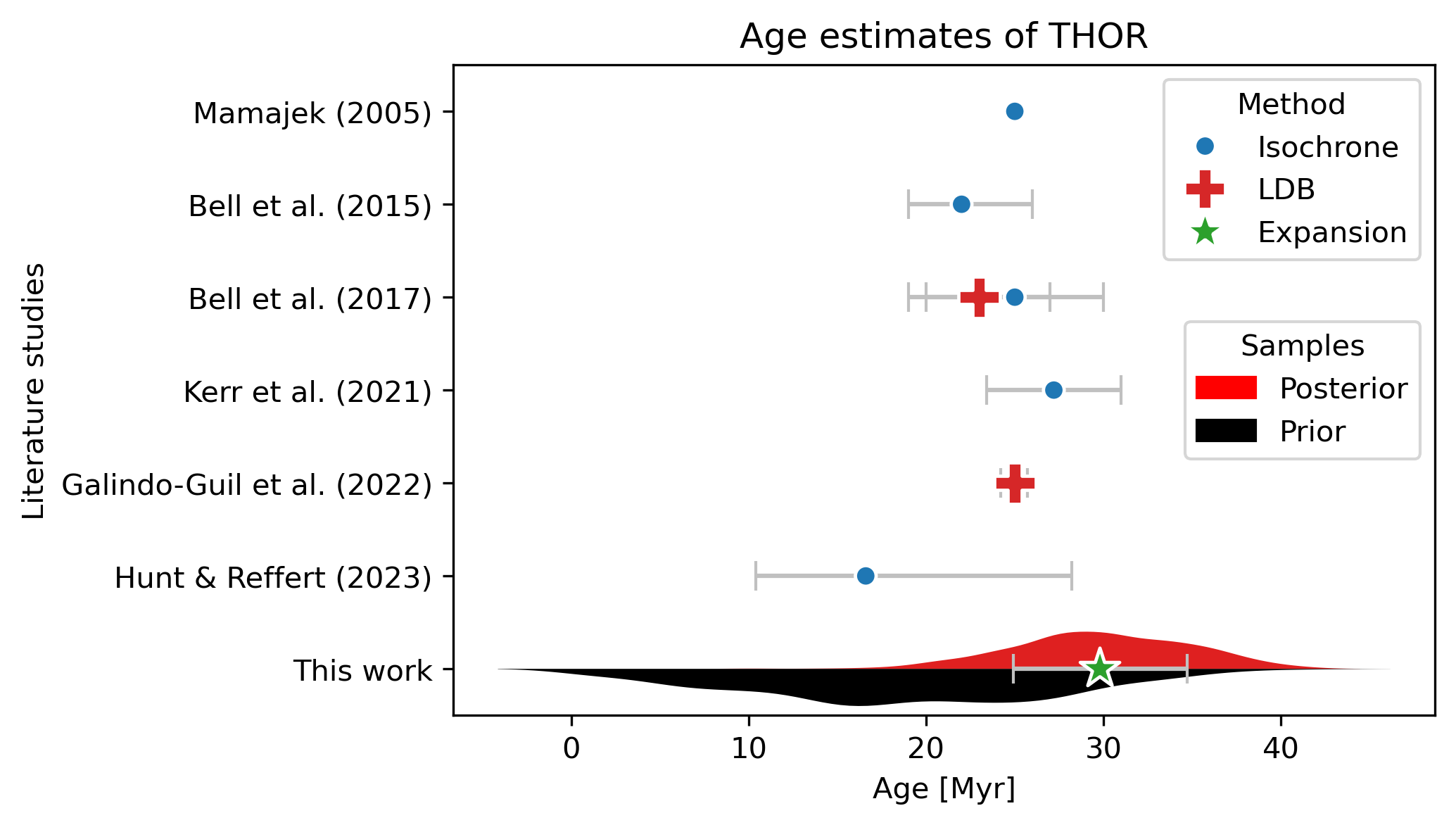}
\caption{Age distribution for THOR. The violin plot shows kernel density estimates obtained from samples of the prior (bottom) and posterior (top) distributions. Previous literature age estimates are included for comparison.}
\label{figure:THOR_lit}
\end{figure}

\begin{figure}[ht!]
\centering
\includegraphics[width=\columnwidth]{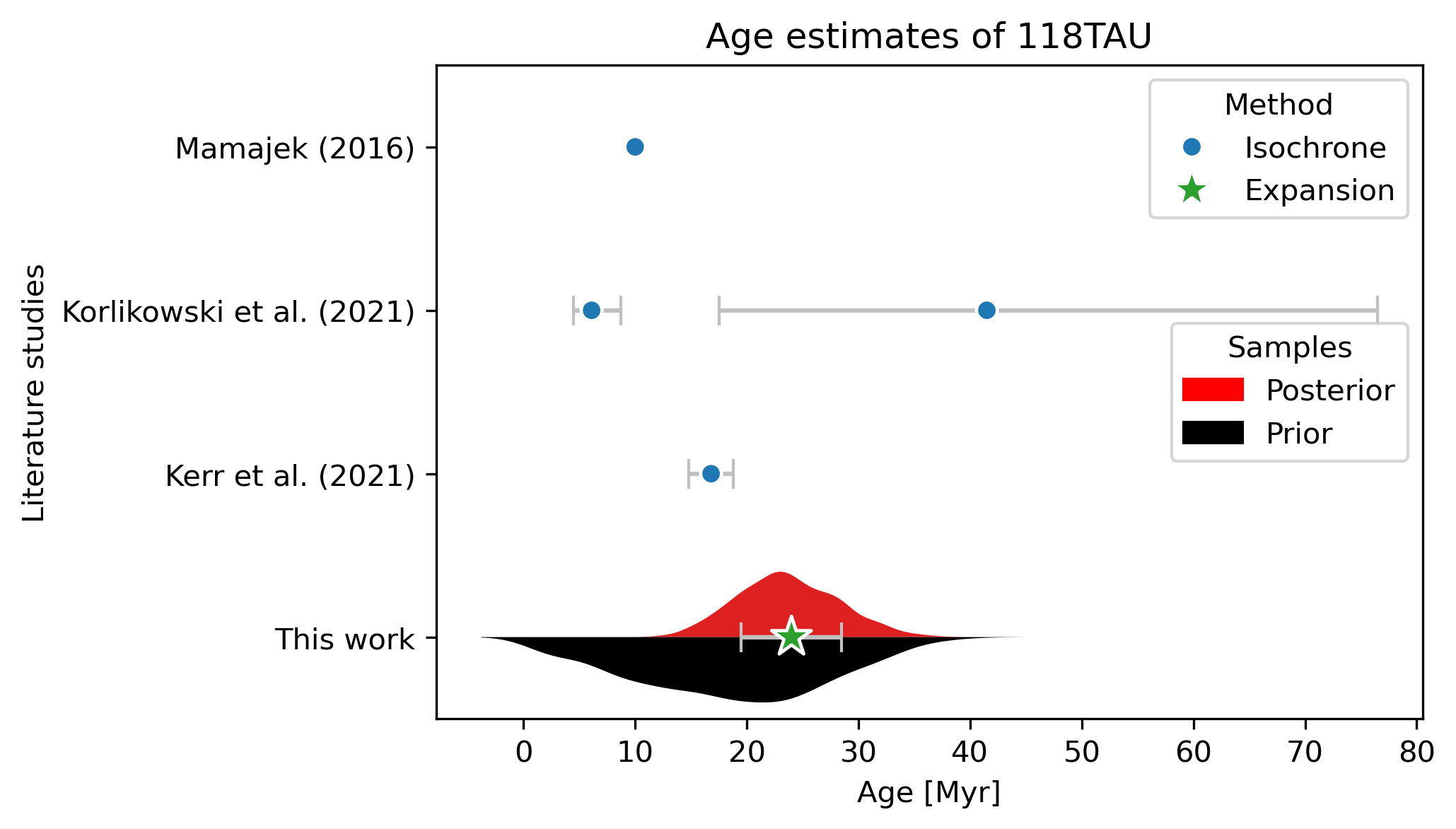}
\caption{Age distribution for 118TAU. The violin plot shows kernel density estimates obtained from samples of the prior (bottom) and posterior (top) distributions. Previous literature age estimates are included for comparison. The two age estimates by \citet{2021AJ....162..110K} correspond to those of their young 118TauE and old 118TauW populations.}
\label{figure:118TAU_lit}
\end{figure}

The age of THOR has been determined in the literature through several methods, but mostly the nuclear ones of isochrone fitting and LDB (see Fig. \ref{figure:THOR_lit}). On the contrary, 118TAU has been poorly characterised, most likely due to its entanglement with THOR \citep{2022AJ....164..151L}, Cas-Tau \citep{2018AJ....156..302D}, Taurion \citep{2015A&A...584A..26B}, and Taurus-Auriga \citep{2017ApJ...838..150K}. Figure \ref{figure:118TAU_lit} shows the few literature age estimates that we were able to gather for this association. 

Recently, \cite{2022AJ....164..151L} attempted to measure the kinematic age of his 32 Orionis association, which includes members of THOR and 118TAU, finding inconsistent expansion age values ranging from 15 Myr to 47 Myr, and a traceback age between 1.5 Myr and 17 Myr. Moreover, he provided a relative isochrone age difference between members of the classical THOR and 118TAU associations of only $3.6\pm2.9$ Myr. We interpret the lack of consistency in the kinematic ages and the apparent coevality of THOR and 118TAU reported by \cite{2022AJ....164..151L} as consequences of the phase-space entanglement of the several populations and substructures within his membership list. 

Concerning THOR's age, our estimate is consistent (at 1$\sigma$) and compatible (at 95\% HDI) with most of the literature estimates obtained through isochrone fitting and LDB, except with the recent estimate by \cite{2023A&A...673A.114H}. Moreover, our age estimate falls within the span of the expansion age reported by \cite{2022AJ....164..151L}. We notice that, despite the young prior ($20\pm10$ Myr) that we chose, our posterior age distribution is older and more compact than the prior, thus indicating that it was informed by the data.

Regarding the age of 118TAU, Fig. \ref{figure:118TAU_lit} shows that our expansion age is older than previous estimates but compatible (within 2$\sigma$) with the most recent ones in the \textit{Gaia} era \citep{2021A&A...650A..48K,2021AJ....162..110K}. We notice that the two age estimates by \citet{2021AJ....162..110K} correspond to the ages of their young 118TauE and old 118TauW substructures, with 6 and 3 members, respectively, which they identified on the basis of their XYZ distribution. However, our phase-space analysis (see Sect. \ref{results:THOR_118TAU}) indicates that given the current data and census of members, there is no evidence to reject the null hypothesis that 118TAU is a single population, and thus contains no substructures. 

To the best of our knowledge, this is the first time that an expansion age is derived for 118TAU \citep[the one derived by][was for his 32 Orionis superstructure]{2022AJ....164..151L}. Nonetheless, studies with other dating techniques, such as traceback or LDB, are still needed to confirm this age.

The expansion ages and close phase-space proximity of THOR, 118TAU, and Háap may suggest that these groups formed out of the same molecular cloud at different star-formation episodes. This hypothesis will require future work and data, particularly, more complete spectroscopic coverage to obtain precise RV and Lithium equivalent width measurements.

\subsection{Háap}
\label{discussion:haap}

\begin{figure}[ht!]
\centering
\includegraphics[width=\columnwidth]{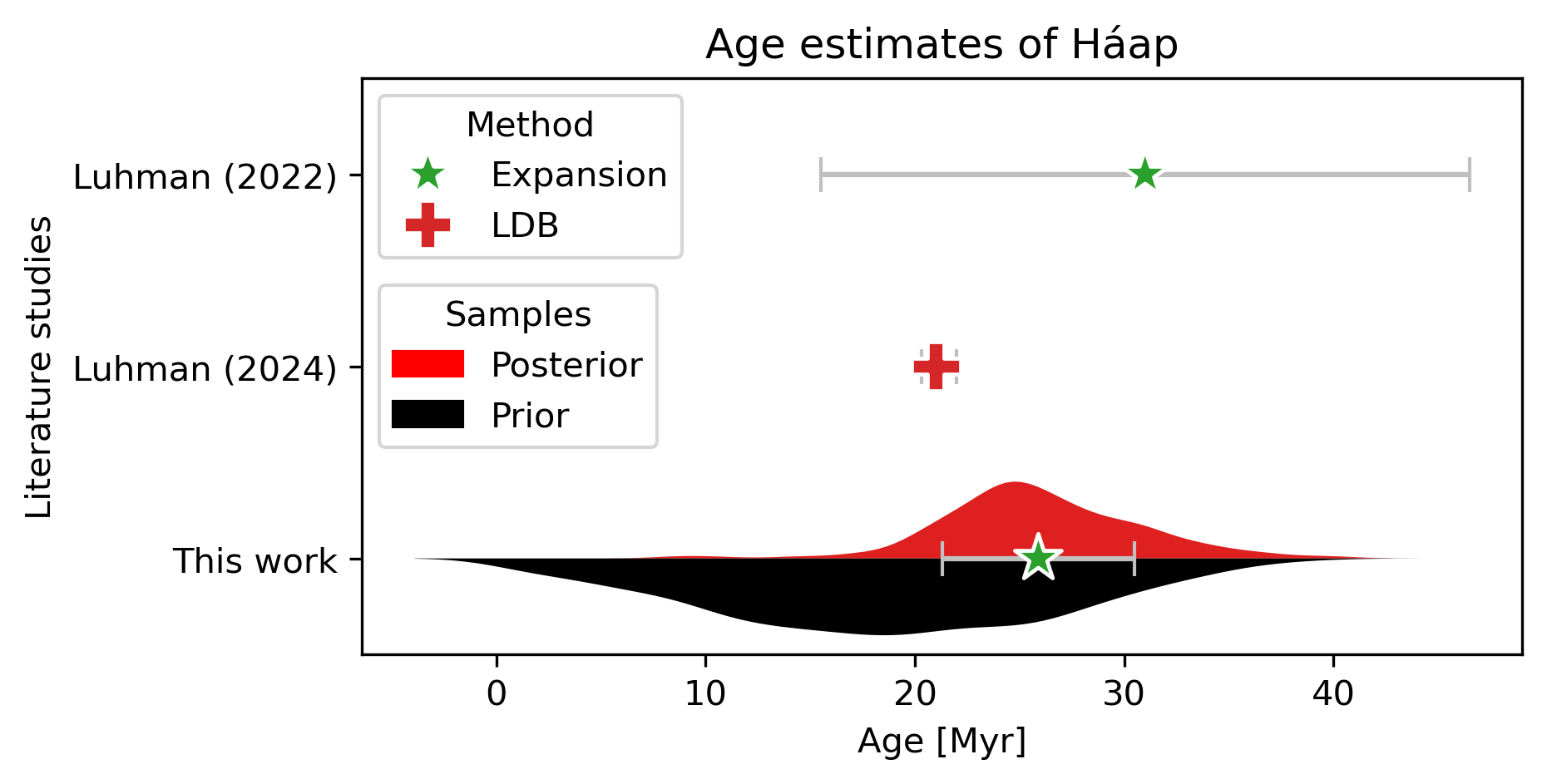}
\caption{Age distribution for Háap. The violin plot shows kernel density estimates obtained from samples of the prior (bottom) and posterior (top) distributions. Previous literature age estimates for 32 Orionis are included for comparison.}
\label{figure:Háap_lit}
\end{figure}

Given that Háap has been proposed for the first time in this work, its age estimate is also the only one we currently have. Nonetheless, Fig. \ref{figure:Háap_lit} shows our expansion age estimate together with the expansion age interval and LDB age derived by \cite{2022AJ....164..151L} and \cite{2024AJ....168..159L}, respectively, for members of his 32 Orionis supergroup. As can be observed, our age estimate is consistent with these two.

In an attempt to compare our age estimate with other ones from the literature, we cross-matched our list of Háap members with the recent \textit{Gaia} DR3 surveys of nearby stellar associations by \citet{2021ApJ...917...23K}, \citet{2022ApJ...939...94M}, and \citet{2024A&A...686A..42H}. With that by \citet{2021ApJ...917...23K}, we found 26 Háap members in common with the Greater Taurus supergroup but only one assigned to 118TAU. With that by \citet{2022ApJ...939...94M}, we found 9 Háap members in common with 118TAU. Finally, with that by \citet{2024A&A...686A..42H}, we only found one cross-match associated with THOR. Therefore, we conclude that the Háap population has been elusive in the recent literature works and was discovered thanks to \citet{2022AJ....164..151L} membership list and our robust statistical methodology.

As discussed in the previous section, Háap's age and phase-space parameters suggest that it may be related to THOR and 118TAU, and thus, to the controversial old population in the Taurus region \citep[see][]{2018AJ....156..271L,2017ApJ...838..150K}. However, future spectroscopic data and traceback studies will be needed to confirm this link.

\subsection{$\beta$-Pictoris}
\label{discussion:BPIC}

\begin{figure}[ht!]
\centering
\includegraphics[width=\columnwidth]{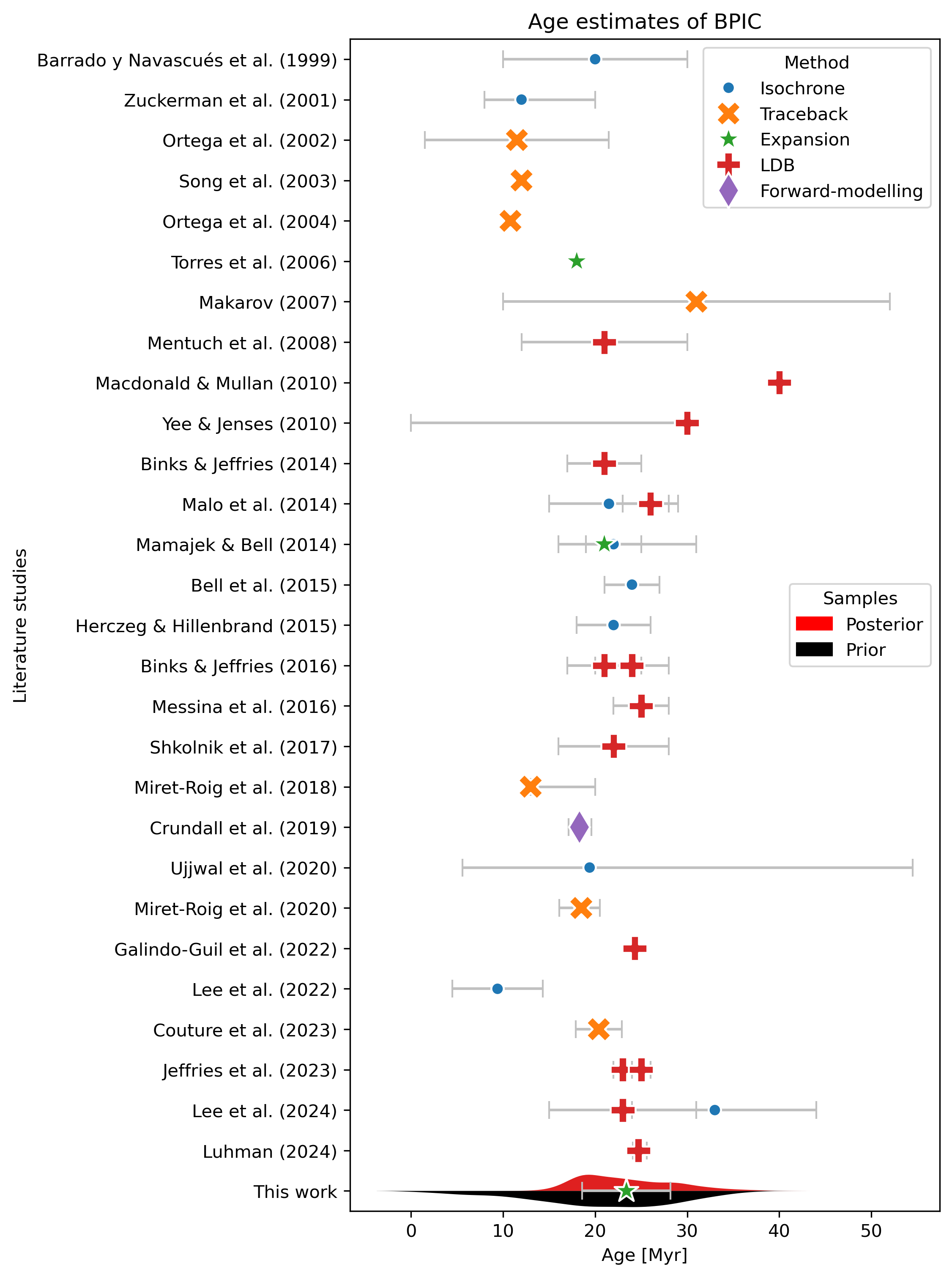}
\caption{Age distribution for BPIC. The violin plot shows kernel density estimates obtained from samples of the prior (bottom) and posterior (top) distributions. Previous literature age estimates are included for comparison.}
\label{figure:BPIC_lit}
\end{figure}

Due to its proximity, young age, and importance for exoplanet searches and star-formation analysis, the age of BPIC has been the focus of several studies. \citet{2024MNRAS.tmp...30L} provide a review of the previous studies and the challenges faced by the various age dating techniques. Comparing our Bayesian expansion age estimate with those from the literature (see Fig. \ref{figure:BPIC_lit}, which is an update of Fig. 1 from \citealt{2024MNRAS.tmp...30L}) we observe that it is compatible with almost all age determination based on the LDB and isochrone fitting methods, particularly with those of recent studies. On the contrary, our expansion age is older and 1$\sigma$ inconsistent (although perfectly compatible within the 95\% HDI) with the trace-back dating method by \citet{2020A&A...642A.179M} and the forward-modelling by \citet{2019MNRAS.489.3625C}.

The main physical difference between the expansion rate method and the trace-back and forward-modelling ones is that the expansion method assumes that the members expand freely, while the other two assume that the members' trajectories are influenced by the Galactic potential. The Galactic potential will effectively accelerate the members and, as a consequence, change the expansion rate of the association as a whole, but it is expected that the accumulated effects of the Galactic potential be important only after 40 Myr (see, for example, \citealt{1952BAN....11..414B}, and the discussion in Sect. 6.3 of \citealt{expansion_method}).

One possible explanation for the age difference between our Bayesian expansion rate method and the dynamical ones from the literature concerns the assumed phase-space geometry of the association. In \citet{2019MNRAS.489.3625C}, the authors assume that, at birth time, the association's size was isotropic, in other words, the authors assume the same standard deviation $\sigma_{XYZ}$ for the three spatial dimension and the same one $\sigma_{UVW}$ for the three velocity dimensions. On the contrary, our method does not assume the phase-space geometry of the association, aside that it follows a Gaussian distribution. In the dynamical trace-back methods by \citet{2023ApJ...946....6C} and \citet{2020A&A...642A.179M}, the age is estimated based on metrics computed from a single dimension of the positional ellipsoid and the arithmetic mean of the three of them, respectively. On the contrary, our method combines the three dimensions of the expansion rate components together with their correlations into a single age estimator \citepalias[see, Sect. 3 of][]{expansion_method}. Interestingly, the age determination that \citet{2023ApJ...946....6C} make based on the determinant of the covariance matrix, which is a metric that robustly estimates the size of the association taking into account not only the dispersions in the three axes but also their correlations, have larger uncertainties than those of the metrics the authors choose to report. For example, our age estimate inferred based on those authors' membership list with their RV data, $26.2\pm5.2$ Myr, is 1$\sigma$ consistent with the $23.1\pm6.3$ Myr that those authors found with the determinant metric. This agreement shows that the complex phase-space geometry of BPIC was captured by the trace-back study of those authors, although they decided to report the age resulting from a single high contrast and low uncertainty metric. Finally, it is also highly likely that a non-isotropic phase-space ellipsoid, as that of BPIC, would affect the different literature metrics in a similar way in which the discrepant values of the $\vec{\kappa}$ components introduce biases in the expansion rate frequentist age estimators \citepalias[see Sect. 6.1 of][]{expansion_method}. Thus, we encourage future users of the trace-back method to test the impact and possible biases of non-isotropic position and velocity ellipsoids on the age determinations resulting from different metrics.

In our phase-space analysis of this association, we identified two populations: the classical BPIC and Balaam (Luhman2024-B). Also, we found two substructures within BPIC (Luhman2024-AA and Luhman2024-AB) that, based on the available data, we were unable to identify as different populations (see Sect. \ref{results:BPIC}). These substructures have expansion ages of $25.7\pm5.5$ Myr (Luhman2024-AA) and $23.7\pm5.4$ Myr (Luhman2024-AB), which are consistent with the rest of BPIC age estimates, and particularly with the $23.4\pm4.8$ inferred based on the membership list by \cite{2020A&A...642A.179M}. Thus, we conclude that our phase-space analysis confirms the hypothesis proposed by \citet{2020A&A...642A.179M} on the presence of two coeval substructures in BPIC. Those authors identified these two kinematic substructures only at the birth time of BPIC and thanks to their detailed trace-back analysis. However, they were unable to disentangle these populations at the present time, most likely due to the low-number statistics of their membership list.

Finally, our Bayesian phase-space analysis allows us to conclude that the discrepancy between BPIC's dynamical and nuclear ages may have its origin in its complex phase-space geometry and substructures. When the dating methods overlook this complexity, by, for example, neglecting statistical correlations, they may result in underestimated uncertainties and apparent age tensions.

\subsection{Balaam}
\label{discussion:Balaam}

\begin{figure}[ht!]
\centering
\includegraphics[width=\columnwidth]{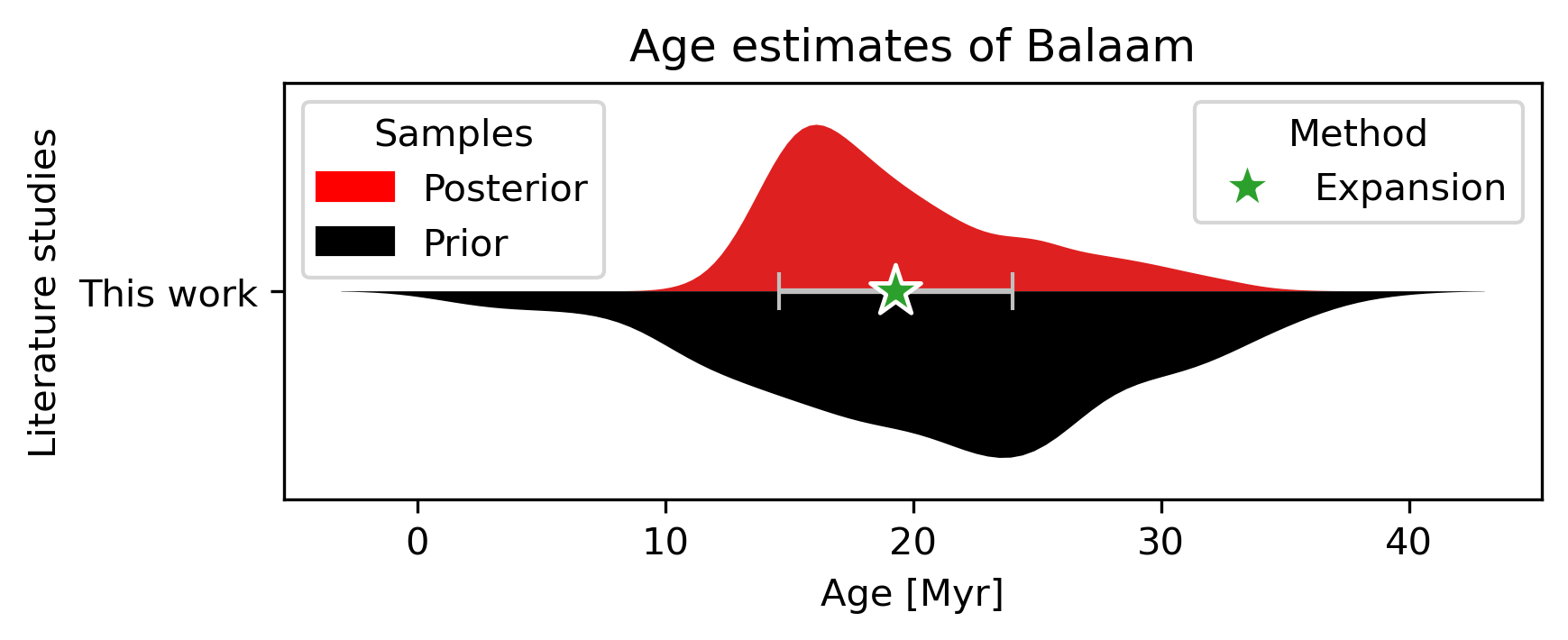}
\caption{Age distribution for Balaam. The violin plot shows kernel density estimates obtained from samples of the prior (bottom) and posterior (top) distributions.}
\label{figure:Balaam_lit}
\end{figure}

Balaam is a new stellar association whose members were previously entangled with those of BPIC in the literature studies. Our phase-space methodology allows us to identify it and provide an age estimate that is younger than that of BPIC. 

Most likely, the discrepant literature age estimates of BPIC may have been biased towards younger ages due to the presence of Balaam. Future work will be needed to both increase its census of members and RV coverage.

\subsection{Octans}
\label{discussion:OCT}

\begin{figure}[ht!]
\centering
\includegraphics[width=\columnwidth]{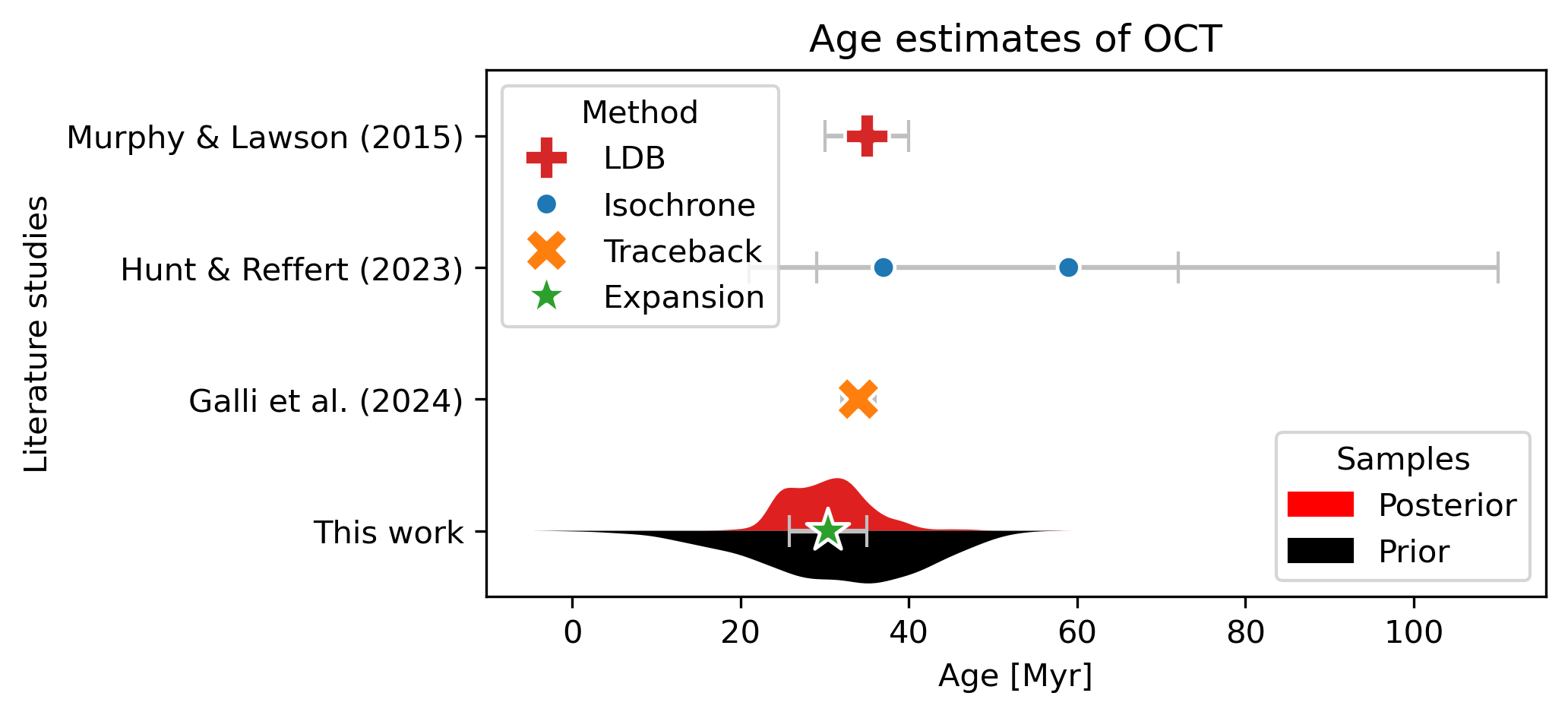}
\caption{Age distribution for OCT. The violin plot shows kernel density estimates obtained from samples of the prior (bottom) and posterior (top) distributions. Previous literature age estimates are included for comparison. The two age estimates by \cite{2023A&A...673A.114H} correspond, in ascending order, to those of their groups CWNU1178 and HSC1998.}
\label{figure:OCT_lit}
\end{figure}
 
In OCT, our methodology was able to identify two partially entangled substructures within the UNION list, UNION-A and UNION-B, with non-negligible weights and Mahalanobis distances that almost suffice to distinguish them as populations. This evidence, in combination with the large physical extent of OCT, which is more than twice that of the rest of the LYSA in the X and Y directions, allows us to hypothesise that this association has been squeezed by tidal forces into this elongated and almost split group of stars. This hypothesis is also supported by the trace-back analysis performed by \citet{2024A&A...689A..11G}, in which those authors find that at the moment of birth, the size of OCT was similar to that of other associations. Another hypothesis is that UNION-A and UNION-B are two distinct populations that our methodology is unable to disentangle given the current data. This hypothesis is supported by the bimodal age distribution of UNION-A. More data is still needed to clarify the current state of OCT.

Comparing our expansion age with those from the literature (see Fig. \ref{figure:OCT_lit}), we observe that it is compatible (within the 95\% HDI) with the isochrone age estimate by \cite{2023A&A...673A.114H} on their group CWNU1178. Moreover, our expansion age is consistent (within 1$\sigma$) with the previous trace back and LDB age estimates by \cite{2024A&A...689A..11G} and \cite{2015MNRAS.447.1267M}, respectively.

\subsection{HSC 2597}
\label{discussion:HSC2597}

\begin{figure}[ht!]
\centering
\includegraphics[width=\columnwidth]{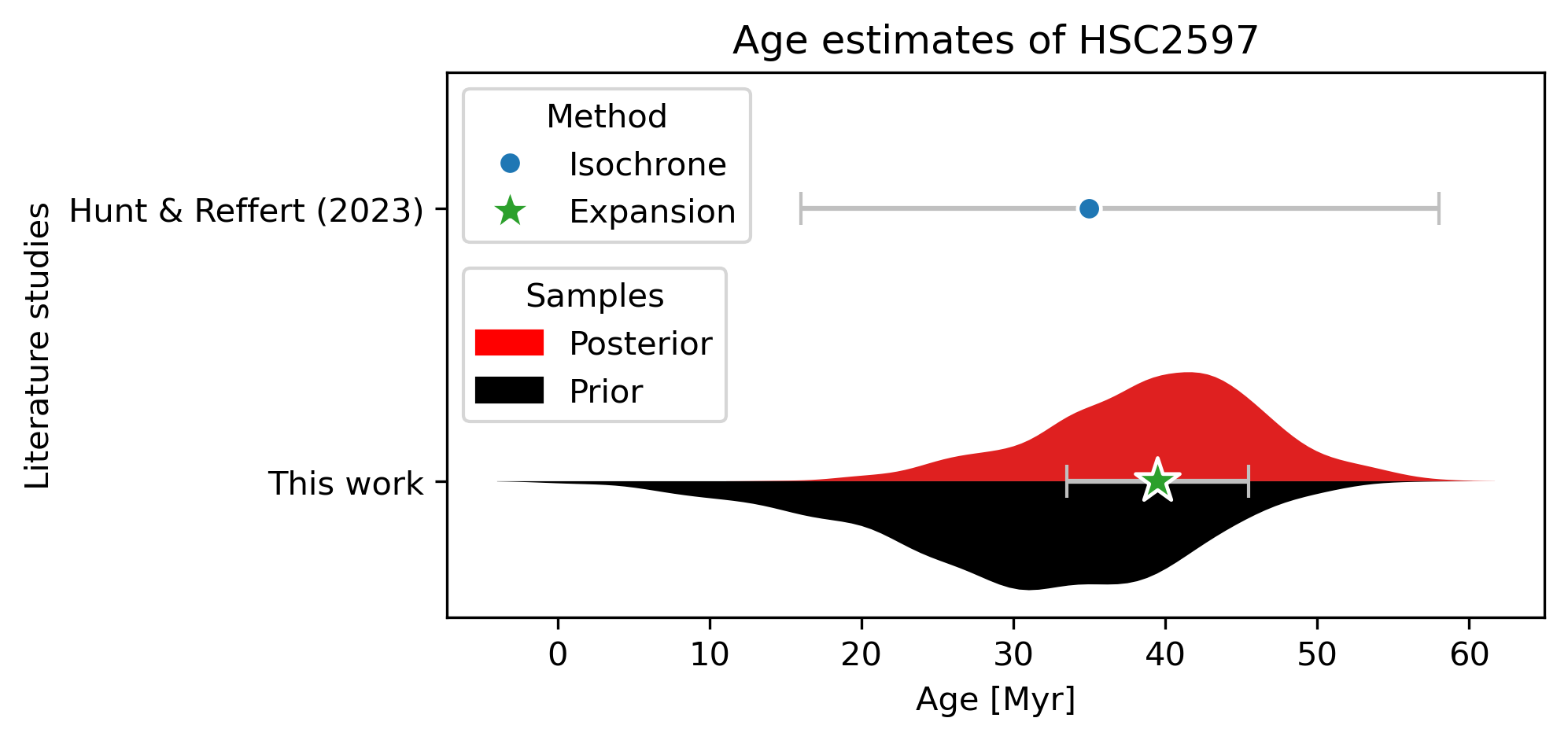}
\caption{Age distribution for HSC 2597. The violin plot shows kernel density estimates obtained from samples of the prior (bottom) and posterior (top) distributions. The isochrone age by \cite{2023A&A...673A.114H} is included for comparison. }
\label{figure:HSC2597_lit}
\end{figure}

Our phase-space methodology allows us to independently discover  \cite{2023A&A...673A.114H} group HSC 2597 within the literature membership lists of OCT.  Our expansion age of this group is not only consistent with that of the previous authors ($35_{-19}^{+23}$) but provides a more precise value, with an uncertainty three times smaller in a population six times smaller. We nonetheless notice that, due to the low number of members, our age estimate may underestimate the true age of the system.

\subsection{Columba}
\label{discussion:COL}
\begin{figure}[ht!]
\centering
\includegraphics[width=\columnwidth]{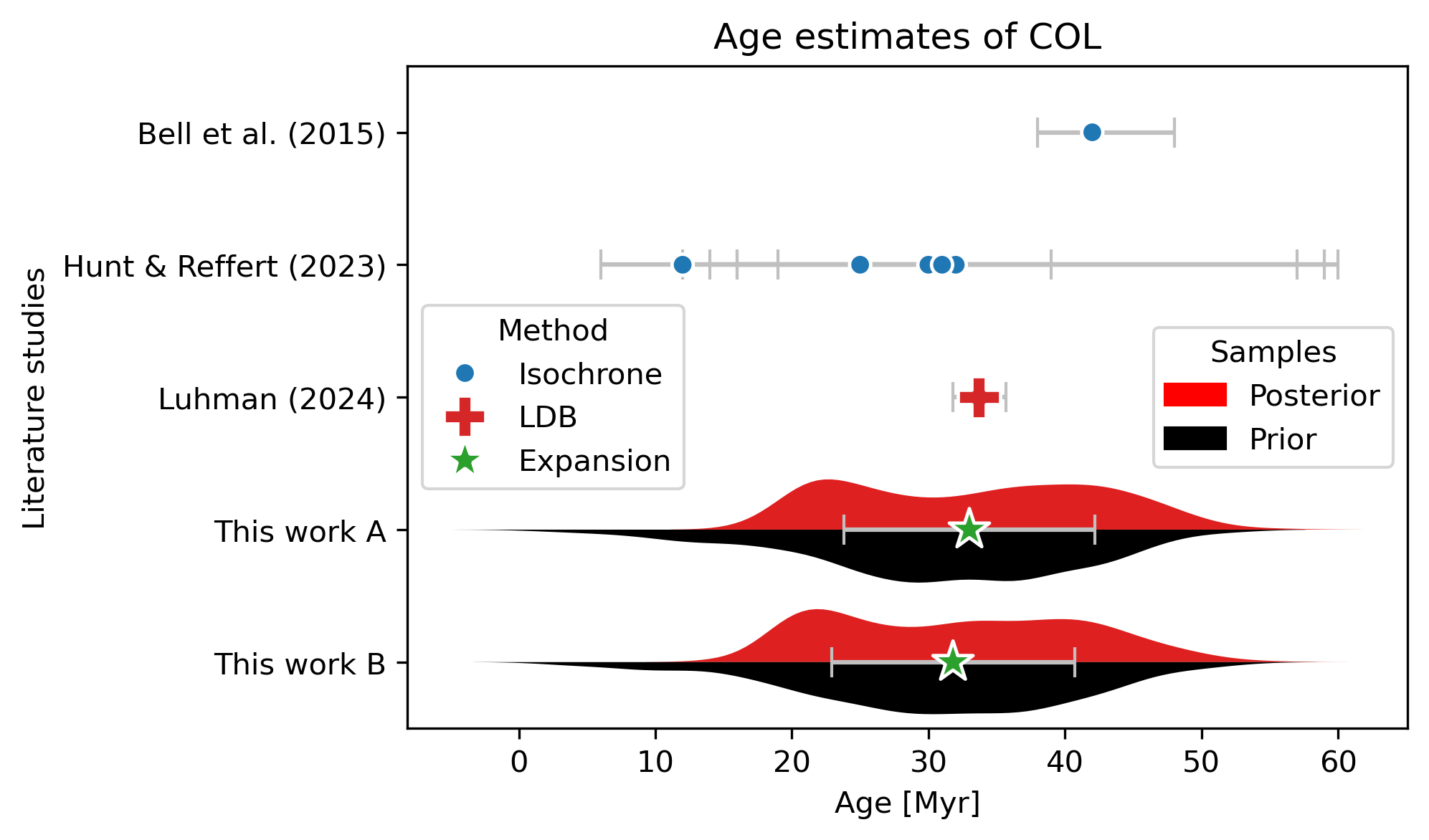}
\caption{Age distribution for COL. The violin plot shows kernel density estimates obtained from samples of the prior (bottom) and posterior (top) distributions. Previous literature age estimates are included for comparison.  The ages by \cite{2023A&A...673A.114H} correspond, in ascending order, to their groups HSC 1900, HSC 1684, HSC1766, Alessi 13, and $\beta$-Tucanae.}
\label{figure:COL_lit}
\end{figure}

As shown in Fig. \ref{figure:COL_lit}, our expansion age is consistent (within 1$\sigma$) with the age estimates by \citet[][$33.7^{+2.0}_{-1.9}$ Myr]{2024AJ....168..159L} with LDB and by \citet[][$45^{+11}_{-7}$ Myr]{2015MNRAS.454..593B} with isochrone. Given that the later authors worked with pre-\textit{Gaia} data, most likely, their much older isochrone age determination was biased by the presence of the older OMAU population (see Sect. \ref{discussion:OMAU}). On the contrary, we found that the COL membership list by \citet{2024AJ....168..159L} has no contaminants, although with only a 50\% RV coverage. 

Our age estimate is also consistent (within 1$\sigma$) with the age estimates by \cite{2023A&A...673A.114H} of their groups  HSC 1684, HSC1766, Alessi 13, and $\beta$-Tucanae but incompatible with HSC 1900. We compare our age estimates with these previous groups because they all show partial entanglement with our groups COL A and COL B. 

Figure \ref{figure:COL_lit} also shows that both COL A and COL B have a bimodal age distribution. This bimodality suggests the presence of substructures that our strict identification criterion may have overlooked. The presence of substructures in both COL A and COL B is a hypothesis which may be supported by the partial entanglement with the groups mentioned above. Future work will be needed to improve the RV of this LYSA and unravel its possible populations and substructures.

\subsection{$\omega$-Aurigae}
\label{discussion:OMAU}
\begin{figure}[ht!]
\centering
\includegraphics[width=\columnwidth]{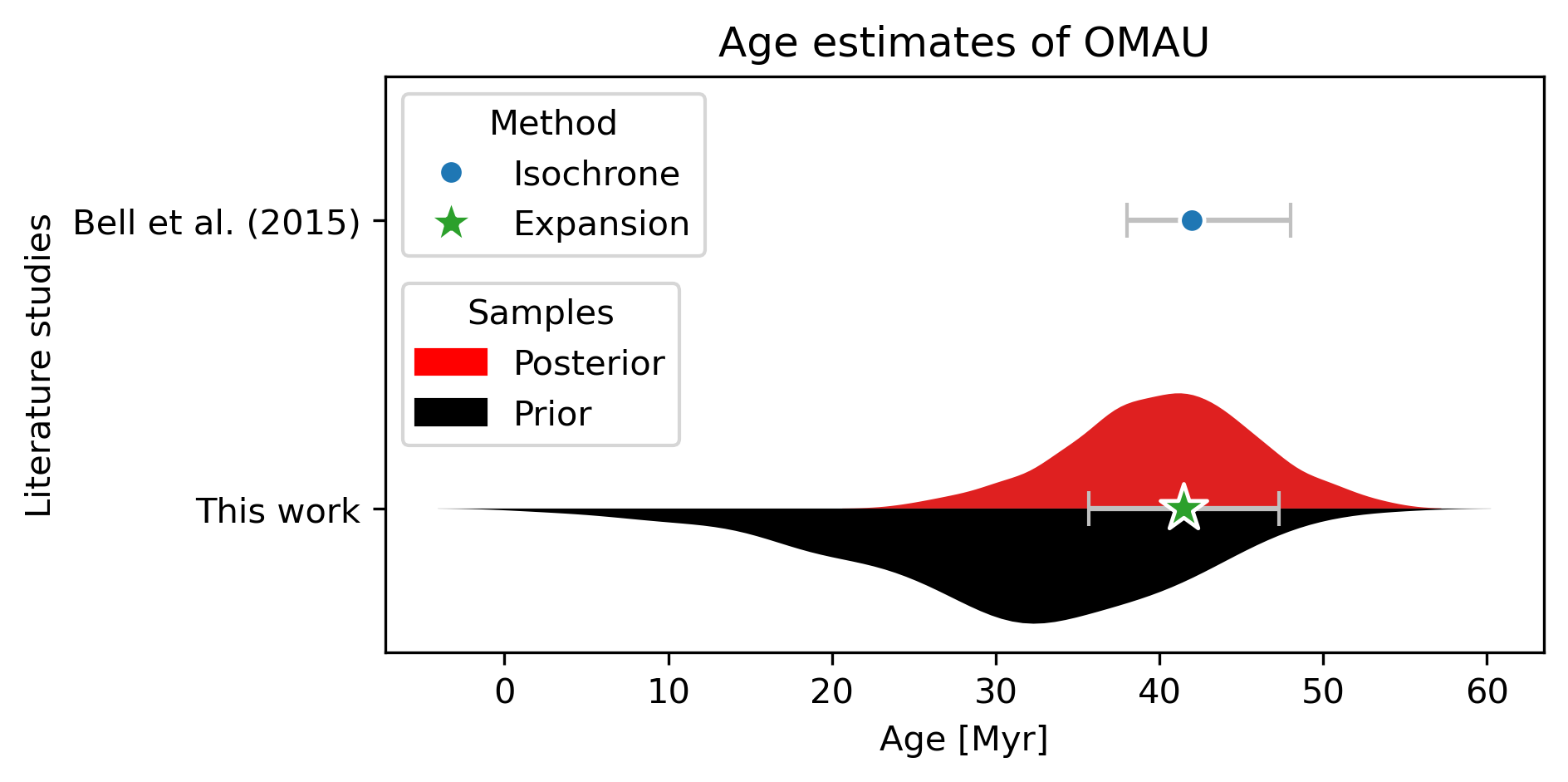}
\caption{Age distribution for OMAU. The violin plot shows kernel density estimates obtained from samples of the prior (bottom) and posterior (top) distributions. The isochrone age estimate by \citet{2015MNRAS.454..593B} for COL is included for comparison.}
\label{figure:OMAU_lit}
\end{figure}

OMAU is a new population whose expansion age indicates an age difference of 7 Myr with COL (based on the UNION-C list). Although this age difference is compatible with the age uncertainties of both populations (6-9 Myr), it points towards a shared origin. As shown in Fig. \ref{figure:OMAU_lit}, our expansion age is consistent (within 1$\sigma$) with the isochrone age of COL by \citet[][$45^{+11}_{-7}$ Myr]{2015MNRAS.454..593B}. As mentioned in the previous section, the study of the later authors was done on pre-\textit{Gaia} membership lists, where the entanglement of COL and OMAU was present.

Due to their similar, although not identical, ages, we hypothesise that COL and OMAU may have formed out of the same parent molecular cloud in two episodes of star formation. Future work will be needed to increase the census and RV coverage of this new association, which becomes, to the best of our knowledge, the nearest association to the Sun, located at only 36 pc and making it an excellent place to search for brown-dwarfs and exoplanets.

\subsection{Carina}
\label{discussion:CAR}

\begin{figure}[ht!]
\centering
\includegraphics[width=\columnwidth]{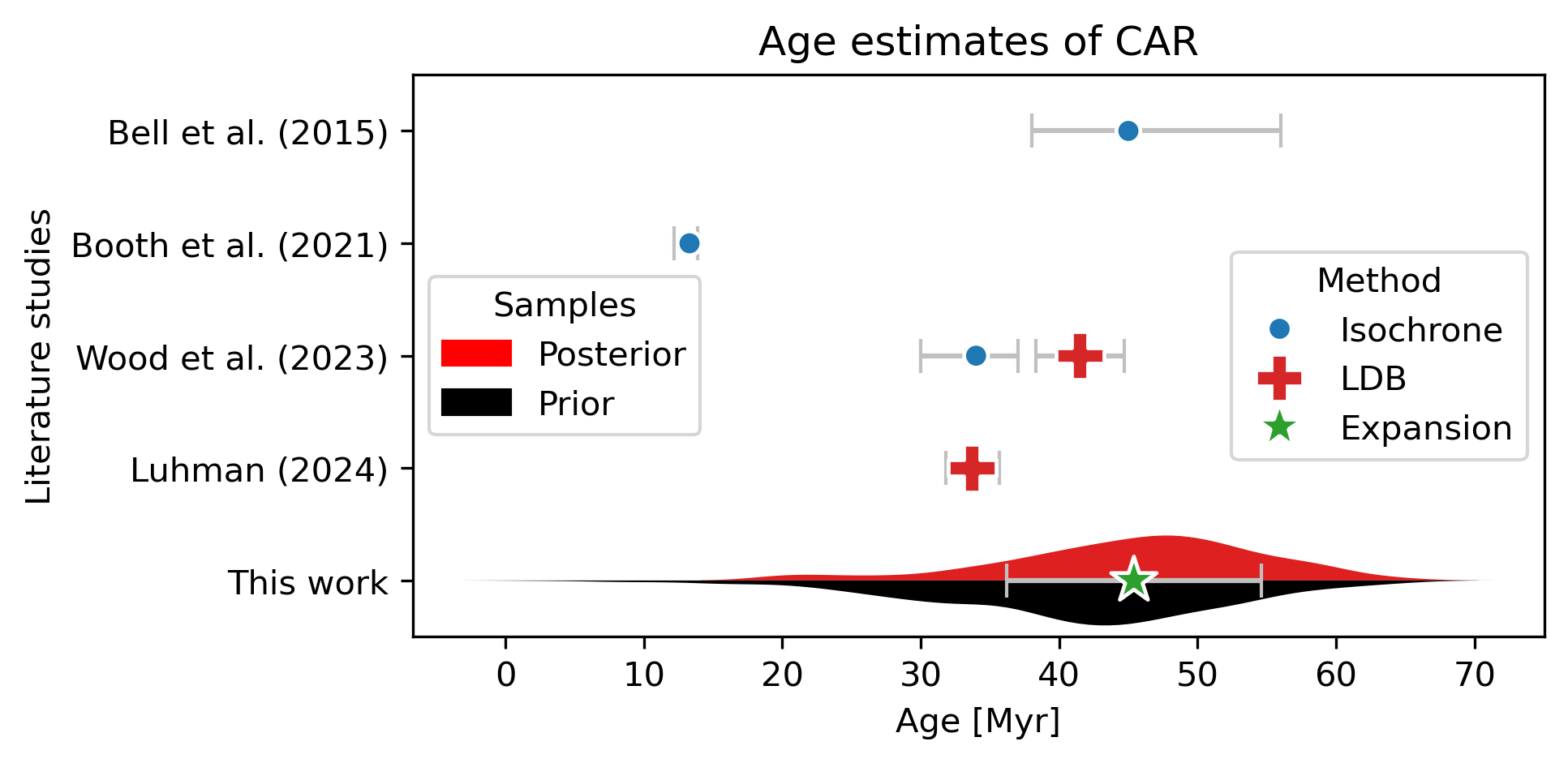}
\caption{Age distribution for CAR. The violin plot shows kernel density estimates obtained from samples of the prior (bottom) and posterior (top) distributions. Previous literature age estimates are included for comparison. }
\label{figure:CAR_lit}
\end{figure}

Figure \ref{figure:CAR_lit} shows that our expansion age for CAR is consistent (within 1$\sigma$) with the LDB ages by \cite{2015MNRAS.454..593B} and \cite{2023AJ....166..247W} and with the isochrone age by the latter authors. However, it is only compatible (within the 95\% HDI) with the LDB age estimate by \cite{2024AJ....168..159L}, although we notice that this latter estimate was obtained for the Car-Ext population. 

It is important to notice that our posterior age distribution resembles that of our prior. Thus indicating that the dataset lacks constraining information to update our prior. For this reason, our CAR age estimate must be taken as a lower limit of the true age of the system (see Sect. \ref{discussion:methodological_caveats}). 

\subsection{Platais 8}
\label{discussion:Platais8}

\begin{figure}[ht!]
\centering
\includegraphics[width=\columnwidth]{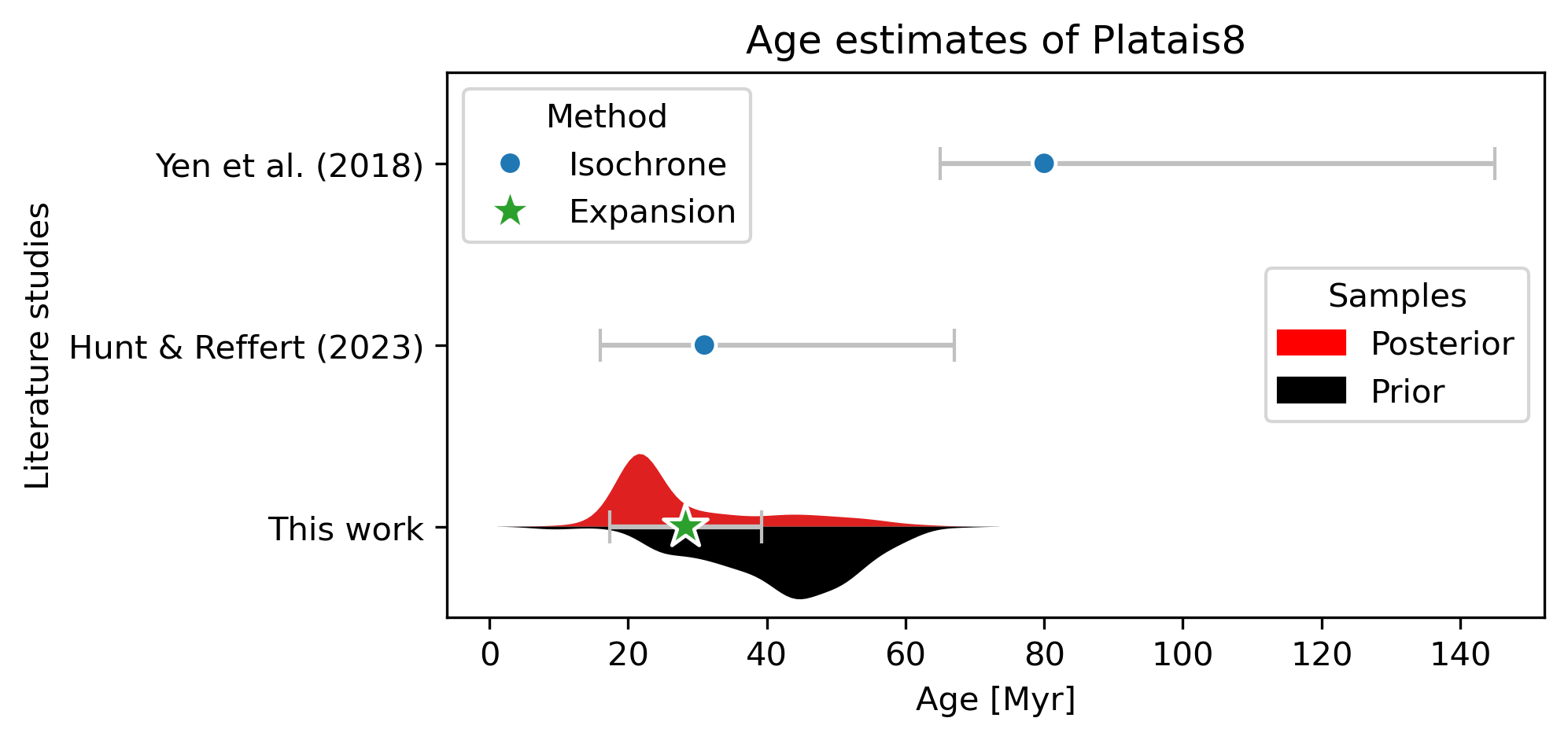}
\caption{Age distribution for Platais 8. The violin plot shows kernel density estimates obtained from samples of the prior (bottom) and posterior (top) distributions. Previous literature age estimates are included for comparison.}
\label{figure:Platais8_lit}
\end{figure}

Platais 8 was found within the Car-Ext list by \cite{2024AJ....168..159L}, which the later author already reports as the most distant cluster within this extended population. Moreover, the latter author reports that Car-Ext is a single large and spatially and kinematically continuous population. However, given our criterion to identify populations (see Sect. \ref{methods:substructures}), we identify several populations within the Car-Ext list. 

As can be seen in Fig. \ref{figure:Platais8_lit}, our age estimate for Platais 8 is consistent (within 1$\sigma$) with the isochrone age by \citet{2023A&A...673A.114H} and incompatible with that by \citet{2018A&A...615A..12Y}. Moreover, the bulk of the age posterior distribution is younger than the bulk of the prior, thus indicating a clear informed posterior. The long and positive posterior tail may be an indication of either some remanent contaminants from nearby associations.

The fact that our expansion age method, given its simple assumptions and limitations, was able to provide an age estimate for young open clusters opens promising venues for the use of this method in other stellar systems in the local neighbourhood. 

\subsection{Nal and Chem}
\label{discussion:NalChem}

\begin{figure}[ht!]
\centering
\includegraphics[width=\columnwidth]{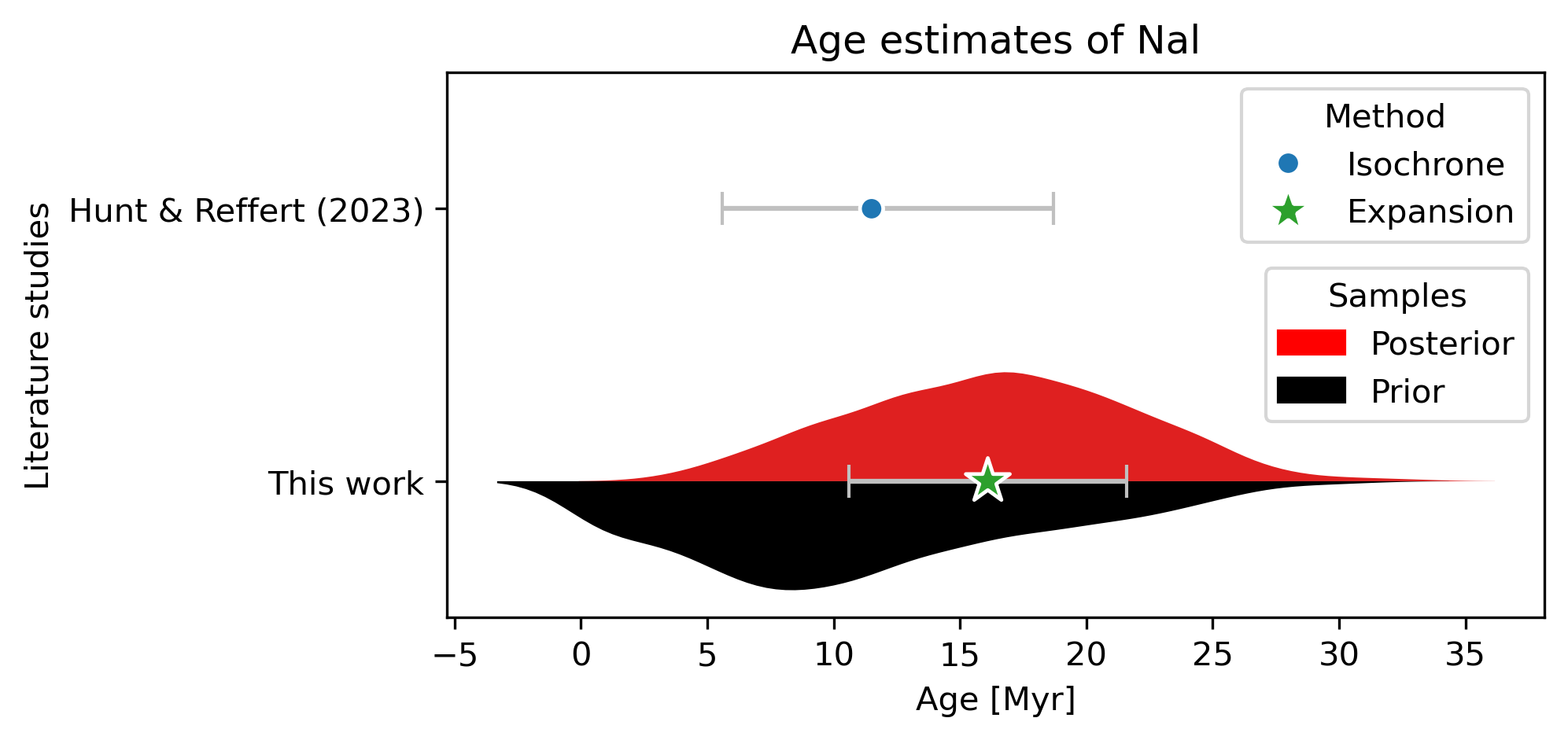}
\caption{Age distribution for Nal. The violin plot shows kernel density estimates obtained from samples of the prior (bottom) and posterior (top) distributions. The previous literature age estimate by \citet{2023A&A...673A.114H} for HSC 2139 is included for comparison.}
\label{figure:Nal_lit}
\end{figure}

\begin{figure}[ht!]
\centering
\includegraphics[width=\columnwidth]{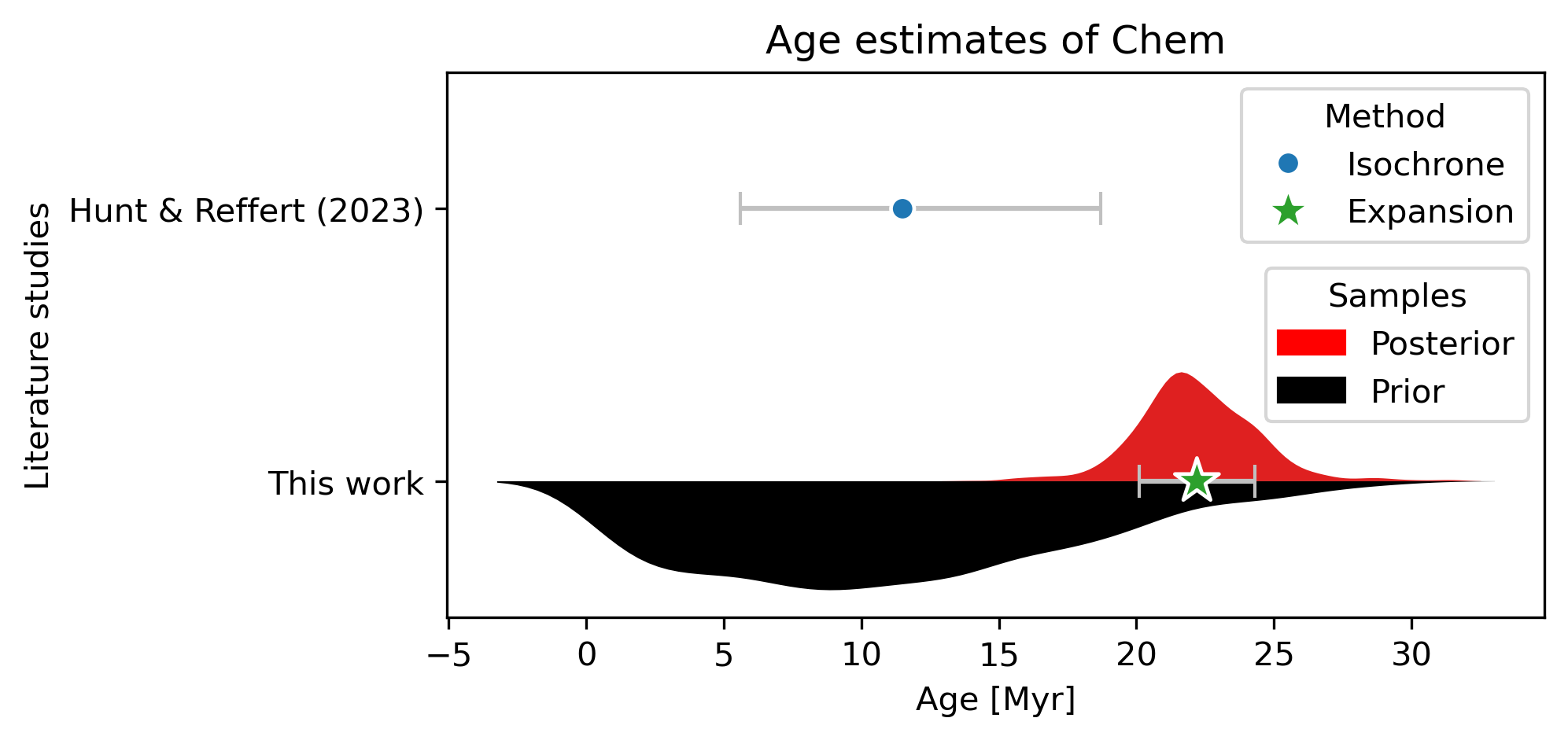}
\caption{Age distribution for Chem. The violin plot shows kernel density estimates obtained from samples of the prior (bottom) and posterior (top) distributions. The previous literature age estimate by \citet{2023A&A...673A.114H} for HSC 2139 is included for comparison.}
\label{figure:Chem_lit}
\end{figure}

The Nal and Chem new associations were discovered within the membership list by \cite{2024AJ....168..159L} thanks to our phase-space methodology. Figures \ref{figure:Nal_lit} and \ref{figure:Chem_lit} show the posterior age estimates of Nal and Chem, respectively, as compared to that of HSC 2139 by \citet{2023A&A...673A.114H}, which, to the best of our knowledge, is the only literature reference that includes both systems. On one hand, the age of Chem is not compatible (within the 95\% HDI) with that of HSC 2139 and shows a shrinkage of the posterior with respect to the prior represented by its narrow uncertainty (2.1 Myr), thus  confirming its independence as a population. On the other hand, the posterior age distribution of Nal is consistent (within 2$\sigma$) with the isochrone age of HSC 2139. Nonetheless, we notice that at the large distance of Nal (184 pc), the low astrometric precision and low RV coverage with only four sources result in this large age uncertainty. On the contrary, the precise age estimate of the Chem association proves that the applicability domain of the Bayesian expansion rate method can be extended up to 170 pc with uncertainties at the 10\% level, provided that the number of sources is larger than a hundred.

Nal and Chem are two new promising associations for testing the validity of star-formation theories at the 150-200 pc domain.  Their age similarity and phase-space proximity indicate that these two systems most likely formed together out of the same parent molecular cloud. Future studies will be needed to both unravel nearby associations formed out of the same parent cloud and to increase the RV coverage, particularly for the Nal association.

\subsection{Tucana-Horologium}
\label{discussion:THA}

\begin{figure}[ht!]
\centering
\includegraphics[width=\columnwidth]{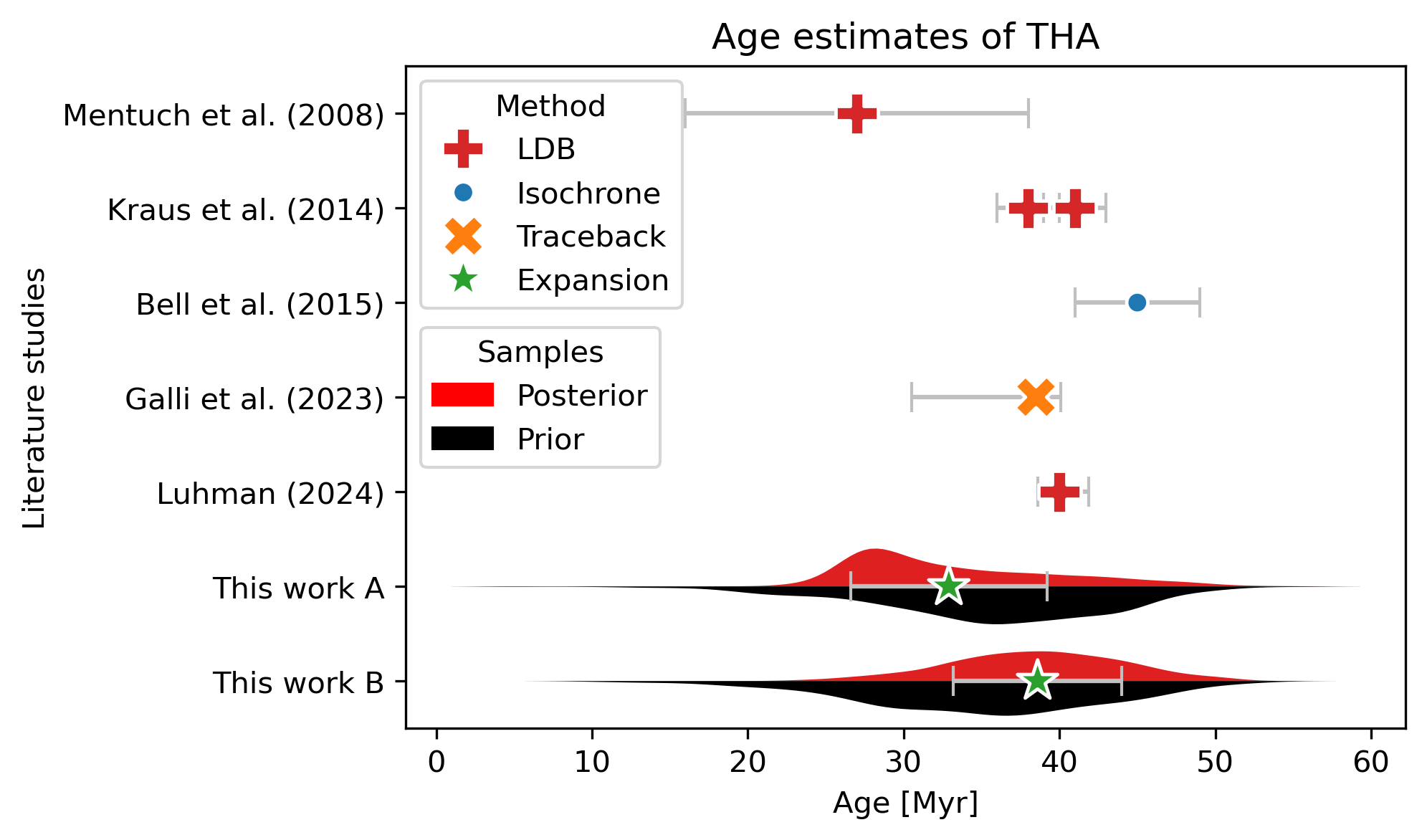}
\caption{Age distribution for THA A and B. The violin plots show kernel density estimates obtained from samples of the prior (bottom) and posterior (top) distributions. Previous literature age estimates are included for comparison.}
\label{figure:THA_lit}
\end{figure}

In THA, our expansion ages for substructures A and B (Lee+2019-A and Lee+2019-B, respectively) are consistent (within 1$\sigma$) with all previous age estimates, except with the isochrone one by \cite{2015MNRAS.454..593B}, with which it is only compatible (within the 95\% HDI). Most likely, the isochrone age estimate of the latter authors was biased towards older ages due to the presence of a nearby old population, such as that of substructure B. 

Concerning the substructures of THA, we observe the following. On the one hand, substructure A (Lee+2019-A) has a posterior age distribution with a peak at 28 Myr and a long positive tail. This peak is consistent with the age derived by \citet[][in their Table 4]{2023MNRAS.520.6245G} for the X direction alone. Thus allowing us to conclude that both dynamical methods can unravel the internal kinematics of THA. On the other hand, substructure B (Lee+2019-B) has a posterior age distribution that is older, but still compatible, with that of substructure A. Therefore, we hypothesise that substructure B is part of a different population that, given the current data, remains entangled with substructure A. 

Future work will be needed to improve the census of members of the substructures of THA and their RV. Moreover, due to the close proximity and relatively old age, the internal kinematics and expansion age of THA will benefit from improving phase-space modelling beyond the linear velocity field.

\subsection{Method caveats}
\label{discussion:methodological_caveats}

We categorise the caveats of our methods into exogenous and endogenous ones. We consider exogenous all those sources affecting the input data, such as astrometric and radial velocity biases, unresolved binaries and contaminants. On the other hand, we consider endogenous effects as those related to the model construction and specification, such as the prior distributions. 

\subsubsection{Exogenous caveats}
\label{discussion:exogenous}

The input data passed to our methods may contain contaminants and unresolved binaries that could bias our estimates. We distinguish between contaminants and binaries because the former, once identified, can be removed from the input list while unresolved binaries are members of the association but with possibly discrepant RV \citep[see Assumption 5 of][]{expansion_method}. One one hand, we expect that contaminants will have a minimum impact in our parameter estimation given the successive filtering stages that we applied, first by removing clearest parallax and RV outliers (see Sect. \ref{methods:outliers}), and then by the iterative phase-space decontamination steps of our method (see Sect. \ref{methods:decontamination}). On the other hand, the impact of unresolved binaries has also been minimised by our filtering criteria, in which the RV of 2$\sigma$ outliers or sources with RUWE>1.4 is masked as missing. We consider these threshold values conservative enough to minimise the biases introduced by unresolved binaries.

The astrometric data that we use can also be biased. However, it comes entirely from the \textit{Gaia} DR3, for which the systematics have been thoroughly analysed in \citet{2021A&A...649A...2L}. The parallax and proper motions angular correlations are corrected as part of the default \textit{Kalkayotl} routines using the recipe corrections provided by \citet{2021A&A...649A...2L}. 

Our RV GLASSS data (see Sect. \ref{data:radial_velocities}) can also be subject to biases. In \citet{2023A&A...674A...5K} and \citet{2023A&A...674A...7B}, the authors have compared the accuracy of \textit{Gaia} DR3 RV with respect to other surveys, including the ones used in the GLASSS data. Although the previous authors derived corrections for the systematics they found, the \textit{Gaia} DR3 archive was not corrected for it, given that, as the previous authors point out, the RV of individual sources can behave differently. Thus, we do not apply those corrections. Concerning the Simbad radial velocities, we query the archive for the RV of \textit{Gaia} DR3 sources within 200 pc from the Sun and found a global shift with respect to \textit{Gaia} DR3 of $0.1\pm3.5\rm{km\,s^{-1}}$, which we deemed statistically insignificant and therefore, we do not correct for it.

\subsubsection{Endogenous caveats}
\label{discussion:endogenous}

We identify as endogenous caveats those related to the phase-space complexity of LYSAs, our criteria to discern populations from substructures, the linear velocity field model, and the selected prior. We now briefly comment on them.

Our methods could be failing at identifying phase-space substructures due to the series of assumptions upon which they were constructed. First, we assume that the phase-space distribution of LYSAs is Gaussian or can be modelled as a mixture of Gaussians \citepalias[see Assumption 4 of][]{expansion_method}. Second, we assume that two populations can be identified as such if 95\% of their phase-space distributions are not overlapped  \citepalias[see Sect. \ref{methods:substructures} and Assumption 4 of][]{expansion_method}. Third, we assume that the velocity space of a LYSA can be described by a linear model. As shown here several times, our methods face convergence issues with the sampling algorithm. In the majority of cases, these resulted from the presence of substructures. However, these failures can also arise from the intrinsic complexity of the velocity field, which is not captured by a simple linear model.

Finally, the prior could also be an important source of bias, particularly when the datasets lack constraining information.  This lack of information generally manifests as an age posterior being similar to the prior, as in the case of CAR (see Fig. \ref{figure:CAR_lit}), where there is only a mild shrinkage of the posterior. In such cases, our age estimates can only be considered as a lower limit because, due to the asymmetry of the transformation \citepalias[see][]{expansion_method}, a low-signal expansion would result in an older age but not in a younger one.

\section{Conclusions}
\label{conclusions}

In this work, we aimed at inferring the expansion ages of classical LYSAs using publicly available data, literature membership lists, and open source codes. In doing so, our phase-space methodology allows us to discover new LYSAs hidden in the literature membership lists and rediscover others recently found thanks to \textit{Gaia} data and improved clustering algorithms \citep[e.g][]{2023A&A...677A..59R}. We independently rediscovered the Musca-Foreground, HSC 2597, and Platais 8 systems and discovered and characterised the following new ones: Háap, Balaam, Nel, Chem, and OMAU. With this last one being the closest stellar association to the Sun.

Our results for the 18 LYSAs represent, to date, the largest and most homogenous compilation of expansion ages and kinematic patterns of stellar associations in the solar neighbourhood. We hope that this compilation serves not only to the study of individual systems, like host-disk stars and exoplanets, but also to improve the formation history of the solar neighbourhood.

Although the methodology that we use here offers improved ways to remove contaminants, minimise the effects of unresolved binaries, identify phase-space substructures and populations, and incorporate the complex phase-space geometry of stellar associations, it has caveats that could be further minimised with improved data, particularly with an increased census of members and RV coverages. Finally, the analysis of stellar associations in the local vicinity will strongly benefit from improved phase-space modelling beyond the simple linear velocity field.

\section*{Data availability}
Tables B.2 and B.3 are only available in electronic form at the CDS via anonymous ftp to cdsarc.u-strasbg.fr (130.79.128.5) or via http://cdsweb.u-strasbg.fr/cgi-bin/qcat?J/A+A/
 
\begin{acknowledgements}
%Referee
We thank the anonymous referee for the useful comments that helped to improve the presentation of these results.
%Javier
JO acknowledge financial support from "Ayudas para contratos postdoctorales de investigación UNED 2021". "La publicación es parte del proyecto PID2022-142707NA-I00, financiado por
MCIN/AEI/10.13039/501100011033/FEDER, UE".
%Phillip's
P.A.B.G. acknowledges financial support from the São Paulo Research Foundation (FAPESP) under grant 2020/12518-8.
%Gaia
This work has made use of data from the European Space Agency (ESA) mission
{\it Gaia} (\url{https://www.cosmos.esa.int/gaia}), processed by the {\it Gaia}
Data Processing and Analysis Consortium (DPAC,
\url{https://www.cosmos.esa.int/web/gaia/dpac/consortium}). Funding for the DPAC
has been provided by national institutions, in particular the institutions
participating in the {\it Gaia} Multilateral Agreement.

% LAMOST
Guoshoujing Telescope (the Large Sky Area Multi-Object Fiber Spectroscopic Telescope LAMOST) is a National Major Scientific Project built by the Chinese Academy of Sciences. Funding for the project has been provided by the National Development and Reform Commission. LAMOST is operated and managed by the National Astronomical Observatories, Chinese Academy of Sciences.

%SoS
Part of this work is based on archival data, software or online services provided by the Space Science Data Center - ASI.

%SDSS
Funding for the Sloan Digital Sky 
Survey IV has been provided by the 
Alfred P. Sloan Foundation, the U.S. 
Department of Energy Office of 
Science, and the Participating 
Institutions. 

SDSS-IV acknowledges support and 
resources from the Center for High 
Performance Computing  at the 
University of Utah. The SDSS 
website is www.sdss4.org.

SDSS-IV is managed by the 
Astrophysical Research Consortium 
for the Participating Institutions 
of the SDSS Collaboration including 
the Brazilian Participation Group, 
the Carnegie Institution for Science, 
Carnegie Mellon University, Center for 
Astrophysics | Harvard \& 
Smithsonian, the Chilean Participation 
Group, the French Participation Group, 
Instituto de Astrof\'isica de 
Canarias, The Johns Hopkins 
University, Kavli Institute for the 
Physics and Mathematics of the 
Universe (IPMU) / University of 
Tokyo, the Korean Participation Group, 
Lawrence Berkeley National Laboratory, 
Leibniz Institut f\"ur Astrophysik 
Potsdam (AIP),  Max-Planck-Institut 
f\"ur Astronomie (MPIA Heidelberg), 
Max-Planck-Institut f\"ur 
Astrophysik (MPA Garching), 
Max-Planck-Institut f\"ur 
Extraterrestrische Physik (MPE), 
National Astronomical Observatories of 
China, New Mexico State University, 
New York University, University of 
Notre Dame, Observat\'ario 
Nacional / MCTI, The Ohio State 
University, Pennsylvania State 
University, Shanghai 
Astronomical Observatory, United 
Kingdom Participation Group, 
Universidad Nacional Aut\'onoma 
de M\'exico, University of Arizona, 
University of Colorado Boulder, 
University of Oxford, University of 
Portsmouth, University of Utah, 
University of Virginia, University 
of Washington, University of 
Wisconsin, Vanderbilt University, 
and Yale University.

%Simbad
This research has made use of the SIMBAD database,
operated at CDS, Strasbourg, France.

%CDS
This research has made use of the VizieR catalogue access tool, CDS, Strasbourg, France.

% Stellarium
This research has made use of the Stellarium planetarium.
% PyMC
We thank the PyMC team for making publicly available this
probabilistic programming language.
\end{acknowledgements}

\bibliographystyle{aa} % style aa.bst
\bibliography{BALYSA} % your references Yourfile.bib

\begin{appendix}
\section{Age prior distributions}
\label{appendix:prior}
In this Appendix, we present the compilation of all literature sources and decisions we took to specify the age prior of each LYSA. We structure the presentation in the same order as in Sect. \ref{results}.

In EPCHA and ETCHA, given their possible common origin, we decided to use the same age prior. For this latter, we set its mode to the 9 Myr, similar to the $8.8^{+2.0}_{-0.4}$ Myr found by \citet{2023A&A...678A..71R} on the base of \textit{Gaia} DR3 and the PARSEC isochrone models (see Sect. 2 of the previous authors). We notice that \citet{2023A&A...678A..71R} report, in their Table 1, similar ages for these two associations, although ETCHA appears slightly but insignificantly older, thus we use the same prior for both associations. We chose a standard deviation of 5 Myr for the age prior to accommodate the age variations found in the literature \citep[see, for example, Table 1 of][]{2018ApJ...860...43G}.

In TWA, we work with an age prior weakly informed by the $10\pm2$ Myr reported by \citet{2023AJ....165..269L}. However, we intentionally provide weak a priori information with an increased age dispersion that accommodates the variation of the literature age estimates. Thus, we set the age prior hyper-parameters to $\mu_\tau\!=\!10$ Myr and $\sigma_\tau\!=\!5$ Myr.

In THOR and 118TAU, we choose to work with an age prior of $20\pm10$ Myr. This prior peaks at the 20 Myr reported by \citet[][in their Table 2]{2019MNRAS.486.3434L} and has a dispersion of 10 Myr, which covers most of the literature age estimates, including the recent age determination by \cite{2022AJ....164..151L} in which the suprastructure of 32 Orionis is $\sim$3 Myr younger than BPIC.

In BPIC, we use as age prior the recent $23\pm8$ Myr age estimate reported by \citet{2024MNRAS.tmp...30L} because it is based on two dating techniques: isochrone fitting and LDB. Moreover, this value is compatible with most of BPIC's literature age estimates, as can be seen in Fig. 1 of the aforementioned authors.

In OCT, we choose to work with a value $34\pm10$ Myr. This prior peaks at the latest literature age estimate of $34\pm2$ Myr by \citet{2024A&A...689A..11G}, although with a dispersion five times larger to provide a weakly informative prior. 

In COL, we work with a prior of $34\pm10$ Myr, which peaks at the latest literature LDB age estimate by \cite{2024AJ....168..159L}. However, to provide a weakly informative age prior, we set the age dispersion to 10 Myr, which is five times larger than the uncertainty reported by the later author.

In CAR, we choose an age prior of $45\pm11$ Myr, which peaks at the isochrone fitting age estimate reported by \cite{2015MNRAS.454..593B}. We selected this literature estimate due to its wide uncertainty and compatibility with the recent LDB age estimate of $41.5\pm3.2$ Myr by \citet{2023AJ....166..247W}. 

In Nal and Chem associations, which were discovered within CAR literature membership lists, we update our age prior to a value of $10\pm10$ Myr. We choose this value following the isochrone age determination of $11_{-6}^{+7}$ Myr done by \citet{2023A&A...673A.114H} for the group HSC 2139, which as explained in Sect. \ref{results:CAR} contains both the Nal and Chem populations.

In THA, we decided to work with an age prior of $38\pm15$ Myr following the age estimate derived by \citet{2023MNRAS.520.6245G}. We increase the age dispersion to accommodate the interval of literature age estimates, and thus, we provide a weakly informative prior.

\section{Additional tables}
\label{appendix:tables}

In Table \ref{table:ages}, we summarise the results of each LYSA with rows corresponding to each literature membership list and the following columns. The first and second columns indicate the association abbreviation and membership list. The third and fourth columns show the RV origin and percentage of coverage (of the final members). The fifth and sixth columns show the percentage of contaminants in the initial list of members and the final number of members, respectively. The seventh column shows the association's distance to the Sun as inferred from the final list of members. The eighth, ninth, and tenth columns show the significant detections (at $2\sigma$ level) of the kinematic patterns of contraction, expansion, and rotation, respectively. Finally, the eleventh and twelfth columns show the 95\% high density interval (HDI) of the age posterior distribution and its mean and standard deviation, respectively. 

Table B.2, available at the CDS, contains statistics of the posterior distribution of linear velocity field parameters as inferred from the various membership lists and RV origins. Columns 1, 2, and 3 list the abbreviated name of the association, the literature membership and RV origin, respectively. Column 4 gives the parameter's statistics (mean, standard deviation and 95\% HDI).  Columns 5 to 10 provide the 6D location in Galactic coordinates. Columns 11 to 16 give the standard deviation of the 6D coordinates (i.e. the association's size). Columns 17 to 34 provide the correlations between pairs of 3D positions and 3D velocities. Column 35 provides the magnitude of the expansion vector $\vec{\kappa}$ while Columns 36 to 38 list the X, Y, and Z components of this vector. Similarly, Column 39 gives the total magnitude of the rotation vector $\vec{\omega}$ while Columns 40 to 42 provide the X, Y, and Z components of this vector. Finally, Columns 43 to 47 give the non-isotropic dilation parameters $W_1$ to $W_5$; for a definition of these parameters, see Eq. 4 of \citet{2000A&A...356.1119L}. 

Table B.3, available at the CDS, contains the final members of each analysed LYSA. Columns 1, 2, and 3 list the abbreviated name of the association, the literature membership and RV origin, respectively. Finally, Column 4 provides the \textit{Gaia} DR3 source id.

\onecolumn
\begin{landscape}
\begin{longtable}{lllccccccccr}
\caption[continued]{\label{table:ages}Members, significant kinematic patterns, and expansion ages.} \\
\toprule
 &  & \multicolumn{2}{c}{RV} & \multicolumn{2}{c}{Members} & Distance & Contraction & Expansion & Rotation & \multicolumn{2}{c}{Age} \\
 &  & origin & coverage & cont. & final &  & [$2\sigma$] & [$2\sigma$] & [$2\sigma$] & HDI & $\mu\pm\sigma$ \\
 &  &  & [\%] & [\%] &  & [pc] &  &  &  & [Myr] & [Myr] \\
Association & Membership &  &  &  &  &  &  &  &  &  &  \\
\midrule
\endfirsthead
\caption[]{Members, significant kinematic patterns, and expansion ages.} \\
\toprule
 &  & \multicolumn{2}{c}{RV} & \multicolumn{2}{c}{Members} & Distance & Contraction & Expansion & Rotation & \multicolumn{2}{c}{Age} \\
 &  & origin & coverage & cont. & final &  & [$2\sigma$] & [$2\sigma$] & [$2\sigma$] & HDI & $\mu\pm\sigma$ \\
 &  &  & [\%] & [\%] &  & [pc] &  &  &  & [Myr] & [Myr] \\
Association & Membership &  &  &  &  &  &  &  &  &  &  \\
\midrule
\endhead
\midrule
\multicolumn{12}{r}{Continued on next page} \\
\midrule
\endfoot
\bottomrule
\endlastfoot
\multirow[t]{7}{*}{EPCHA} & LACEwING & GLASSS & 63.3 & 36.2 & 30 & 101 &  & $\kappa_z$ &  & (6.6,15.0) & $10.5\pm2.2$ \\
 & BANYAN & GLASSS & 51.5 & 15.4 & 33 & 100 &  & $\kappa_z$ &  & (7.2,13.3) & $9.8\pm1.6$ \\
 & SPYGLASS & GLASSS & 64.7 & 0.0 & 17 & 101 &  &  &  & (4.5,14.2) & $9.2\pm2.6$ \\
 & Qin+2023 & GLASSS & 69.7 & 0.0 & 33 & 101 &  & $\kappa_z$ &  & (4.9,13.0) & $8.5\pm2.2$ \\
 & SigMA & GLASSS & 37.8 & 5.1 & 37 & 102 &  & $\kappa_x$,$\kappa_z$ &  & (3.5,10.9) & $6.4\pm2.1$ \\
 & UNION & GLASSS & 50.8 & 26.7 & 63 & 101 &  & $\kappa_z$ &  & (6.1,11.2) & $8.5\pm1.4$ \\
 & UNION-B & GLASSS & 52.0 & 3.8 & 25 & 102 &  & $\kappa_z$ &  & (4.1,15.2) & $9.5\pm3.0$ \\
MuscaFG & UNION-A & GLASSS & 53.8 & 0.0 & 39 & 101 &  & $\kappa_z$ &  & (8.1,13.4) & $10.4\pm1.4$ \\
\multirow[t]{6}{*}{ETCHA} & LACEwING & GLASSS & 55.6 & 21.7 & 18 & 98 &  &  &  & (1.8,14.4) & $8.1\pm3.6$ \\
 & BANYAN & GLASSS & 57.1 & 6.7 & 14 & 98 &  &  &  & (1.1,14.5) & $7.6\pm3.9$ \\
 & SPYGLASS & GLASSS & 60.0 & 11.8 & 15 & 97 &  &  &  & (2.4,13.4) & $6.9\pm3.1$ \\
 & SigMA & GLASSS & 60.0 & 33.3 & 20 & 98 &  &  &  & (2.1,13.9) & $7.6\pm3.4$ \\
 & Hunt+2024 & GLASSS & 61.1 & 5.3 & 18 & 97 &  &  &  & (2.0,12.4) & $6.2\pm3.0$ \\
 & UNION & GLASSS & 60.0 & 41.2 & 20 & 98 &  &  &  & (2.3,14.2) & $7.6\pm3.4$ \\
\multirow[t]{13}{*}{TWA} & LACEwING & GLASSS & 56.2 & 43.9 & 32 & 60 &  & $\kappa_x$,$\kappa_y$ &  & (10.4,15.2) & $12.8\pm1.2$ \\
 & BANYAN & GLASSS & 56.7 & 25.0 & 30 & 65 &  & $\kappa_x$,$\kappa_y$,$\kappa_z$ &  & (9.7,12.6) & $11.1\pm0.8$ \\
 & Lee+2019 & GLASSS & 65.8 & 7.3 & 38 & 59 &  & $\kappa_x$,$\kappa_z$ & $\omega_z$ & (6.6,12.9) & $9.7\pm1.6$ \\
 & Luhman2023 & GLASSS & 61.4 & 14.9 & 57 & 63 &  & $\kappa_x$,$\kappa_y$,$\kappa_z$ &  & (7.5,10.7) & $9.0\pm0.9$ \\
 & Luhman2023 & original & 62.3 & 9.0 & 61 & 63 &  & $\kappa_x$,$\kappa_y$,$\kappa_z$ &  & (7.9,10.9) & $9.2\pm0.8$ \\
 & Miret-Roig+2025 & original & 33.9 & 11.9 & 59 & 63 &  & $\kappa_x$,$\kappa_y$,$\kappa_z$ &  & (7.9,11.3) & $9.5\pm0.9$ \\
 & UNION & GLASSS & 57.1 & 39.8 & 56 & 66 &  & $\kappa_x$,$\kappa_y$,$\kappa_z$ &  & (9.3,12.6) & $10.8\pm0.9$ \\
 & Miret-Roig+2025a & original & 32.4 & 17.8 & 37 & 72 &  & $\kappa_x$,$\kappa_y$,$\kappa_z$ &  & (8.1,12.6) & $10.1\pm1.2$ \\
 & Miret-Roig+2025b & original & 42.1 & 9.5 & 19 & 46 &  &  &  & (9.1,18.2) & $13.5\pm2.4$ \\
 & L23-A & GLASSS & 58.5 & 2.4 & 41 & 63 &  & $\kappa_x$,$\kappa_y$,$\kappa_z$ &  & (8.2,12.2) & $10.2\pm1.0$ \\
 & L23-B & GLASSS & 53.3 & 0.0 & 15 & 65 &  & $\kappa_x$,$\kappa_z$ &  & (6.2,11.1) & $8.5\pm1.3$ \\
 & UNION-A & GLASSS & 51.1 & 2.1 & 47 & 66 &  & $\kappa_x$,$\kappa_y$,$\kappa_z$ &  & (10.0,12.9) & $11.4\pm0.8$ \\
 & UNION-B & GLASSS & 37.5 & 0.0 & 8 & 71 &  &  &  & (7.3,15.8) & $11.4\pm2.2$ \\
\multirow[t]{5}{*}{118TAU} & BANYAN & GLASSS & 72.7 & 8.3 & 11 & 102 &  &  &  & (5.2,30.8) & $17.6\pm7.4$ \\
 & SPYGLASS & GLASSS & 43.3 & 11.8 & 30 & 107 &  &  &  & (16.5,36.2) & $26.2\pm5.1$ \\
 & Moranta+2022 & GLASSS & 84.6 & 27.8 & 13 & 95 &  & $\kappa_y$ &  & (20.7,39.4) & $30.1\pm4.7$ \\
 & UNION & GLASSS & 64.0 & 12.3 & 50 & 103 &  & $\kappa_y$ &  & (26.1,41.0) & $33.5\pm3.9$ \\
 & UNIONS+L22-A & GLASSS & 60.0 & 0.0 & 35 & 107 &  & $\kappa_z$ &  & (15.4,32.4) & $24.0\pm4.5$ \\
\multirow[t]{8}{*}{THOR} & LACEwING & GLASSS & 76.9 & 40.9 & 13 & 97 &  & $\kappa_z$ & $\omega_x$ & (9.7,31.2) & $19.9\pm5.8$ \\
 & BANYAN & GLASSS & 46.2 & 40.9 & 26 & 100 &  &  &  & (18.0,37.0) & $27.1\pm4.9$ \\
 & Lee+2019 & GLASSS & 66.7 & 2.9 & 33 & 102 &  &  &  & (15.4,35.3) & $25.4\pm5.1$ \\
 & SPYGLASS & GLASSS & 62.5 & 27.3 & 8 & 101 &  &  &  & (16.2,37.0) & $26.7\pm5.4$ \\
 & Zerjal+2023 & GLASSS & 57.7 & 0.0 & 26 & 98 &  &  &  & (8.2,31.2) & $18.9\pm6.4$ \\
 & Hunt+2024 & GLASSS & 57.6 & 0.0 & 33 & 99 &  &  &  & (15.6,35.6) & $25.7\pm5.2$ \\
 & UNION & GLASSS & 63.3 & 31.0 & 49 & 100 &  &  &  & (17.9,36.8) & $27.3\pm4.9$ \\
 & UNIONS+L22-B & GLASSS & 66.7 & 0.0 & 45 & 101 &  &  &  & (20.3,39.1) & $29.8\pm4.9$ \\
\multirow[t]{11}{*}{THOR+118TAU} & Luhman2022 & original & 61.5 & 12.4 & 148 & 97 &  & $\kappa_x$,$\kappa_y$,$\kappa_z$ & $\omega_x$,$\omega_y$,$\omega_z$ & (20.5,40.4) & $31.4\pm5.0$ \\
 & Luhman2022 & GLASSS & 62.1 & 17.2 & 140 & 97 &  & $\kappa_x$,$\kappa_y$,$\kappa_z$ & $\omega_x$,$\omega_y$,$\omega_z$ & (19.2,40.4) & $32.1\pm5.4$ \\
 & UNIONS-L22 & GLASSS & 60.7 & 27.0 & 89 & 101 &  & $\kappa_y$,$\kappa_z$ &  & (22.0,36.9) & $29.6\pm3.8$ \\
 & UNIONS+L22 & GLASSS & 61.8 & 25.8 & 144 & 97 &  & $\kappa_x$,$\kappa_y$,$\kappa_z$ & $\omega_x$,$\omega_y$,$\omega_z$ & (22.9,42.0) & $33.4\pm4.8$ \\
 & L22-A & original & 76.7 & 0.0 & 86 & 103 &  & $\kappa_x$,$\kappa_y$,$\kappa_z$ &  & (22.9,37.2) & $29.8\pm3.7$ \\
 & L22-B & original & 43.5 & 0.0 & 62 & 91 &  & $\kappa_x$,$\kappa_y$ &  & (16.1,39.5) & $28.8\pm5.6$ \\
 & L22-A & GLASSS & 67.1 & 1.2 & 85 & 103 &  &  &  & (25.0,40.3) & $32.4\pm4.0$ \\
 & L22-B & GLASSS & 54.4 & 8.1 & 57 & 91 &  & $\kappa_x$,$\kappa_y$ &  & (17.5,34.2) & $24.7\pm4.3$ \\
 & UNIONS+L22-A & GLASSS & 60.0 & 0.0 & 35 & 107 &  & $\kappa_z$ &  & (15.4,32.4) & $24.0\pm4.5$ \\
 & UNIONS+L22-B & GLASSS & 66.7 & 0.0 & 45 & 101 &  &  &  & (20.3,39.1) & $29.8\pm4.9$ \\
 & UNIONS+L22-C & GLASSS & 55.6 & 1.6 & 63 & 90 &  & $\kappa_x$,$\kappa_y$,$\kappa_z$ &  & (17.8,35.9) & $25.9\pm4.6$ \\
\multirow[t]{3}{*}{Háap} & L22-B & original & 43.5 & 0.0 & 62 & 91 &  & $\kappa_x$,$\kappa_y$ &  & (16.1,39.5) & $28.8\pm5.6$ \\
 & L22-B & GLASSS & 54.4 & 8.1 & 57 & 91 &  & $\kappa_x$,$\kappa_y$ &  & (17.5,34.2) & $24.7\pm4.3$ \\
 & UNIONS+L22-C & GLASSS & 55.6 & 1.6 & 63 & 90 &  & $\kappa_x$,$\kappa_y$,$\kappa_z$ &  & (17.8,35.9) & $25.9\pm4.6$ \\
\multirow[t]{13}{*}{BPIC} & LACEwING & GLASSS & 63.2 & 44.1 & 38 & 22 &  & $\kappa_x$ &  & (15.6,33.8) & $24.1\pm5.0$ \\
 & BANYAN & GLASSS & 66.3 & 18.9 & 86 & 31 &  & $\kappa_x$,$\kappa_y$ & $\omega_y$ & (15.4,34.2) & $25.2\pm5.3$ \\
 & Lee+2019 & GLASSS & 61.2 & 26.4 & 103 & 31 &  & $\kappa_x$,$\kappa_y$ & $\omega_x$ & (15.9,36.3) & $26.7\pm6.1$ \\
 & Crundall+2019 & GLASSS & 73.8 & 8.7 & 42 & 34 &  & $\kappa_x$,$\kappa_y$ & $\omega_x$ & (15.6,32.7) & $23.8\pm4.8$ \\
 & Miret-Roig+2020 & GLASSS & 73.1 & 0.0 & 26 & 42 & $\kappa_z$ & $\kappa_x$,$\kappa_y$ &  & (15.1,33.1) & $24.0\pm4.9$ \\
 & Miret-Roig+2020 & original & 73.1 & 0.0 & 26 & 42 &  & $\kappa_x$,$\kappa_y$ & $\omega_y$ & (15.8,33.0) & $23.4\pm4.8$ \\
 & SPYGLASS & GLASSS & 52.6 & 20.8 & 19 & 55 & $\kappa_z$ &  &  & (18.2,37.9) & $28.5\pm5.0$ \\
 & Moranta+2022 & GLASSS & 95.2 & 8.7 & 21 & 44 &  & $\kappa_x$ & $\omega_x$ & (13.7,36.6) & $26.8\pm5.9$ \\
 & Couture+2023 & GLASSS & 87.0 & 8.0 & 23 & 33 &  & $\kappa_x$,$\kappa_y$ & $\omega_x$ & (16.0,35.0) & $26.1\pm5.3$ \\
 & Couture+2023 & original & 87.5 & 4.0 & 24 & 32 &  & $\kappa_x$,$\kappa_y$ & $\omega_x$,$\omega_z$ & (17.4,35.3) & $26.2\pm5.2$ \\
 & Luhman2024 & GLASSS & 60.0 & 14.1 & 165 & 42 &  & $\kappa_x$,$\kappa_y$ & $\omega_x$ & (16.3,33.7) & $26.0\pm4.7$ \\
 & Luhman2024-AA & original & 50.0 & 0.0 & 82 & 37 &  & $\kappa_x$,$\kappa_y$ & $\omega_x$,$\omega_y$ & (15.8,35.4) & $25.7\pm5.5$ \\
 & Luhman2024-AB & original & 68.6 & 0.0 & 35 & 22 & $\kappa_z$ & $\kappa_x$,$\kappa_y$ & $\omega_x$ & (14.7,32.5) & $23.7\pm5.4$ \\
Balaam & Luhman2024-B & original & 47.6 & 0.0 & 63 & 63 &  & $\kappa_x$,$\kappa_y$,$\kappa_z$ & $\omega_x$ & (12.2,29.2) & $19.3\pm4.7$ \\
\multirow[t]{6}{*}{OCT} & LACEwING & GLASSS & 45.5 & 42.9 & 44 & 115 &  & $\kappa_x$,$\kappa_y$ & $\omega_y$,$\omega_z$ & (25.3,43.7) & $33.7\pm4.8$ \\
 & BANYAN & GLASSS & 66.1 & 45.1 & 56 & 134 & $\kappa_z$ & $\kappa_x$,$\kappa_y$ &  & (22.8,40.0) & $30.2\pm4.7$ \\
 & Moranta+2022 & GLASSS & 87.5 & 7.7 & 72 & 156 & $\kappa_z$ & $\kappa_x$,$\kappa_y$ & $\omega_y$,$\omega_z$ & (24.4,42.6) & $33.0\pm4.9$ \\
 & Galli+2024 & GLASSS & 100.0 & 6.9 & 27 & 145 &  & $\kappa_x$,$\kappa_y$ &  & (21.9,38.0) & $29.0\pm4.3$ \\
 & Galli+2024 & original & 100.0 & 6.9 & 27 & 146 &  & $\kappa_x$,$\kappa_y$ &  & (22.8,44.5) & $33.0\pm5.9$ \\
 & UNION-A & GLASSS & 81.2 & 0.0 & 69 & 150 & $\kappa_z$ & $\kappa_x$,$\kappa_y$ & $\omega_z$ & (22.5,39.1) & $30.4\pm4.6$ \\
HSC2597 & UNION-C & GLASSS & 41.2 & 0.0 & 17 & 138 &  &  &  & (25.9,52.7) & $39.6\pm6.9$ \\
\multirow[t]{8}{*}{COL} & LACEwING & GLASSS & 65.2 & 37.1 & 66 & 60 &  & $\kappa_x$ & $\omega_y$ & (21.2,48.2) & $35.1\pm7.2$ \\
 & Lee+2019 & GLASSS & 65.9 & 14.0 & 129 & 70 & $\kappa_z$ & $\kappa_x$,$\kappa_y$ & $\omega_z$ & (18.5,48.3) & $35.5\pm8.3$ \\
 & Moranta+2022 & GLASSS & 100.0 & 0.0 & 11 & 72 &  & $\kappa_y$ &  & (22.0,46.6) & $34.0\pm6.6$ \\
 & Luhman2024 & original & 41.2 & 0.0 & 51 & 58 & $\kappa_z$ & $\kappa_x$ &  & (17.4,45.4) & $31.0\pm7.8$ \\
 & Luhman2024 & GLASSS & 52.9 & 0.0 & 51 & 58 &  & $\kappa_x$ &  & (18.2,43.9) & $30.9\pm6.9$ \\
 & BANYAN-A & GLASSS & 64.3 & 2.3 & 84 & 69 & $\kappa_z$ & $\kappa_x$ & $\omega_x$ & (16.5,47.4) & $31.2\pm8.8$ \\
 & UNION-A & GLASSS & 59.2 & 0.8 & 125 & 71 & $\kappa_z$ & $\kappa_x$,$\kappa_y$ & $\omega_z$ & (18.7,48.7) & $33.0\pm9.2$ \\
 & UNION-B & GLASSS & 56.5 & 0.0 & 46 & 58 & $\kappa_z$ & $\kappa_x$ & $\omega_x$,$\omega_y$ & (18.0,47.2) & $31.8\pm8.9$ \\
\multirow[t]{2}{*}{OMAU} & BANYAN-B & GLASSS & 46.7 & 0.0 & 15 & 31 &  &  & $\omega_y$ & (24.7,49.1) & $37.8\pm6.3$ \\
 & UNION-C & GLASSS & 46.7 & 0.0 & 15 & 36 &  &  &  & (29.3,52.3) & $40.1\pm5.9$ \\
\multirow[t]{9}{*}{CAR} & LACEwING & GLASSS & 66.7 & 22.7 & 51 & 94 & $\kappa_y$,$\kappa_z$ &  &  & (28.3,62.6) & $45.6\pm8.8$ \\
 & BANYAN & GLASSS & 53.0 & 4.6 & 83 & 73 & $\kappa_z$ & $\kappa_x$ & $\omega_x$ & (20.5,56.9) & $38.8\pm10.1$ \\
 & Lee+2019 & GLASSS & 64.4 & 13.2 & 59 & 99 & $\kappa_z$ & $\kappa_x$,$\kappa_y$ & $\omega_x$ & (24.6,56.4) & $39.6\pm8.4$ \\
 & Booth+2020 & GLASSS & 63.6 & 15.4 & 22 & 72 & $\kappa_z$ & $\kappa_x$ &  & (26.3,59.5) & $42.7\pm8.8$ \\
 & UNION & GLASSS & 58.1 & 29.5 & 129 & 83 & $\kappa_z$ & $\kappa_x$ & $\omega_x$,$\omega_z$ & (21.0,61.8) & $44.4\pm11.3$ \\
 & Luhman2024 & GLASSS & 40.3 & 3.9 & 812 & 132 &  & $\kappa_x$,$\kappa_y$ & $\omega_x$,$\omega_y$ & (23.3,55.7) & $41.5\pm8.0$ \\
 & Luhman2024-A & GLASSS & 42.7 & 2.5 & 626 & 123 & $\kappa_z$ & $\kappa_x$,$\kappa_y$ & $\omega_x$,$\omega_y$,$\omega_z$ & (28.0,59.7) & $42.7\pm8.1$ \\
 & Luhman2024-B & GLASSS & 25.1 & 1.8 & 167 & 171 &  & $\kappa_x$ &  & (21.1,60.7) & $44.3\pm11.0$ \\
 & Luhman2024-AB & GLASSS & 44.8 & 0.8 & 259 & 112 & $\kappa_z$ & $\kappa_x$,$\kappa_y$ & $\omega_x$,$\omega_z$ & (24.8,63.3) & $45.4\pm9.2$ \\
Platais8 & Luhman2024-AA & GLASSS & 42.0 & 0.8 & 362 & 133 & $\kappa_z$ & $\kappa_x$,$\kappa_y$ & $\omega_z$ & (16.5,54.4) & $28.3\pm10.9$ \\
Nal & Luhman2024-BA & GLASSS & 11.4 & 0.0 & 35 & 183 & $\kappa_x$,$\kappa_z$ &  &  & (5.8,26.4) & $16.1\pm5.5$ \\
Chem & Luhman2024-BB & GLASSS & 30.3 & 0.0 & 132 & 168 &  & $\kappa_x$ &  & (18.0,26.5) & $22.2\pm2.1$ \\
\multirow[t]{8}{*}{THA} & BANYAN & GLASSS & 63.5 & 3.4 & 85 & 44 & $\kappa_z$ & $\kappa_x$ &  & (24.5,49.4) & $36.1\pm7.3$ \\
 & Lee+2019 & GLASSS & 56.0 & 10.7 & 200 & 43 &  & $\kappa_x$,$\kappa_y$ &  & (24.5,49.8) & $39.5\pm6.7$ \\
 & Galli+2023 & original & 95.2 & 0.0 & 21 & 42 &  & $\kappa_x$ &  & (21.1,47.0) & $32.2\pm8.0$ \\
 & Galli+2023 & GLASSS & 94.7 & 9.5 & 19 & 41 &  & $\kappa_x$ &  & (23.1,46.7) & $32.1\pm6.5$ \\
 & BANYAN-A & GLASSS & 69.1 & 2.9 & 68 & 44 & $\kappa_z$ & $\kappa_x$ &  & (24.9,47.9) & $34.1\pm6.6$ \\
 & BANYAN-B & GLASSS & 35.7 & 6.7 & 14 & 49 &  &  &  & (17.1,46.8) & $32.6\pm7.9$ \\
 & Lee+2019-A & GLASSS & 58.4 & 2.4 & 161 & 42 & $\kappa_z$ & $\kappa_x$ & $\omega_z$ & (24.7,46.9) & $32.9\pm6.3$ \\
 & Lee+2019-B & GLASSS & 54.1 & 0.0 & 37 & 43 &  & $\kappa_x$ &  & (27.9,49.1) & $38.6\pm5.4$ \\
\end{longtable}
\end{landscape}
\FloatBarrier %\usepackage{placeins}
\clearpage

\end{appendix}

\end{document}